\documentclass[structabstract]{aa}
\usepackage{graphicx}
\usepackage{natbib}
\usepackage{amssymb}
\usepackage{amsmath}
\usepackage{array,arydshln}
\newcolumntype{D}{!{\vrule width 1.5 pt}}

\def\arcsec{\hbox{$^{\prime\prime}$}}
\usepackage{txfonts}
\begin{document}
   \title{Astrometry with the MCAO instrument MAD}

   \subtitle{An analysis of single-epoch data obtained in the layer-oriented
            mode\thanks{Based on observations collected at the European Southern
            Observatory, Paranal, Chile, as part of the MAD Guaranteed Time Observations}}

   \author{E. Meyer
          \inst{1,2}
          \and
          M. K\"urster \inst{1}
      \and
      C. Arcidiacono \inst{3,4}
% \inst{3}
      \and
      R. Ragazzoni \inst{3}
      \and
      H.-W. Rix \inst{1}
          }

   \institute{Max Planck Institute for Astronomy (MPIA),
              K\"onigstuhl 17, 69117 Heidelberg, Germany
              \and
	     Leiden Observatory, Leiden University, P.O. Box 9513, 2300 RA
	     Leiden, The Netherlands\\
	     \email{meyer@strw.leidenuniv.nl}
         \and
	     INAF Osservatorio Astronomico di Padova, Vicolo dell'Osservatorio,
	     5, 35122 Padova, Italy
     \and
         INAF Osservatorio Astrofisico di Arcetri, Largo Enrico Fermi, 5, 50125,
Firenze, Italy\\
             }

  \date{Received November 3, 2010; accepted May 11, 2011}

% \abstract{}{}{}{}{}
% 5 {} token are mandatory

  \abstract
%   % {} leave it empty if necessary
     {Current instrument developments at the largest telescopes worldwide
     have provisions for Multi-Conjugated Adaptive Optics (MCAO) modules.
     The large field of view and more uniform correction provided by these
     systems is not only
     highly beneficial for photometric studies but also for astrometric
     analysis of, e.g., large dense clusters and exoplanet detection and characterization.
     The Multi-conjugated Adaptive optics Demonstrator (MAD) is the
     first such instrument and was temporarily installed and tested
     at the ESO/VLT in 2007. We analyzed two globular cluster data
     sets in terms of achievable astrometric precision. Data were
     obtained in the layer-oriented correction mode, one in full MCAO
     correction mode with two layers corrected (NGC~6388) and the other applying ground-layer correction only
     (47~Tuc).}
%   aims heading (mandatory)
     {We aim at analyzing the first available MCAO imaging
     data in the layer-oriented mode obtained with the MAD instrument in terms of astrometric
     precision and stability.}
% %   % methods heading (mandatory)
     {We calculated Strehl maps for each frame in both data sets.
     Distortion corrections were performed and the astrometric
     precision was analyzed by calculating mean stellar positions over
     all frames and by investigation of the positional residuals present
     in each frame after transformation to a master-coordinate-frame.}
     {The mean positional precision for stars between $K = 14-18$~mag
     is $\approx 1.2$~mas in the full MCAO correction mode data of the
     cluster NGC~6388. The precision measured in the GLAO data (47~Tuc)
     reaches $\approx 1.0$~mas for stars corresponding to 2MASS $K$
     magnitudes between 9 and 12. The observations were such that
     stars in these magnitude ranges correspond to the same detector
     flux range. The jitter movement used to scan a
     larger field of view introduced additional distortions in
     the frames, leading to a degradation of the achievable precision.}
     {}

   \keywords{technique: image processing -- instrumentation: adaptive optics, MCAO -- methods:
astrometry -- globular cluster: individual: NGC~6388, 47~Tuc
               }

   \maketitle
%
%________________________________________________________________
\
\section{Introduction}
In classical adaptive optics correction, with one reference star,
the field of view (FoV) is limited by the effect of
anisoplanatism, as only the integrated phase error over the column
above the telescope in the direction to the guide star is measured
and corrected. Turbulence outside this column, e.g. in the
direction of the target, if it cannot be used itself as guide star,
is not mapped and the correction degrades rapidly with growing
separation to the guide star. The average wavefront phase error is
limited to $< 1$\,rad only within the so-called isoplanatic angle,
which is for typical astronomical sites $10\arcsec-20\arcsec$ in
the $Ks$-band and only $3\arcsec$ in the visible. In the case of a
laser guide star as reference source, the phase error is even larger,
due to the low focussing altitude and the resulting
cone-effect \citep{Tallon1990,Yan2005}. Multi Conjugated Adaptive Optics
\citep[MCAO;][]{Beckers1988, Ellerbroeck1994} is an approach to
achieve diffraction limited image quality over larger FoVs of up
to 2-4 arcminutes and hence overcome anisoplanatism. Moderate
averaged Strehl-ratios, in the range of 10\% to 25\%, can be achieved, but with a
higher uniformity of the Point-Spread-Function (PSF) over the FoV.
This is desired for resolving structures of extended sources, such
as galaxies or cores of star clusters. In MCAO the 3-dimensional
structure of the turbulence is reconstructed by means of the
information coming from several guide stars, natural or laser.
Instead of correcting the turbulence integrated over a single
direction, turbulence from different layers is corrected with
several deformable mirrors conjugated to these layers. The maximum
achievable performance on the single reference stars is not as good as
with classical adaptive optics (AO), due to the turbulence above
and below the single corrected layers, but instead the correction
is more uniform over a significantly larger FoV. Typically two
layers are being corrected, the ground layer close to the
telescope and a higher layer at around 8 - 10 km height (dependent
on the site). Most of the turbulence in the atmosphere is
generated in the ground layer. Correcting only this layer
(GLAO=Ground-Layer Adaptive Optics), one can remove the major
contributor to the phase aberrations of the incoming wavefronts
\citep{Rigaut2002}. Two different approaches are in use to combine the
signals from the different reference stars, the Star Oriented (SO)
and the Layer Oriented (LO). In the SO mode,
each reference star is observed by one wavefront sensor (WFS) and
one detector. The information from the different directions of the
guide stars is combined to generate 3D information of the
atmosphere within the mapped FoV. In this approach of turbulence
tomography \citep{Tallon1990} the influence of a single layer can
be computed and corrected with one deformable mirror conjugated to
this layer. The first verification of this approach was done in an
open loop measurement at the Telescopio Nazionale di Galileo (TNG)
\citep{Ragazzoni2000}. In the layer oriented approach
\citep{Ragazzoni2000b}, each WFS and detector is conjugated to one
layer in the atmosphere instead to a single star. The light of
several guide stars is optically co-added to increase the signal-to-noise ratio
(SNR) on
the detector, such that also fainter stars can be used as guide
stars. This increases the sky coverage, the fraction of regions on
the sky that can offer a suitable natural asterism, essentially
for this approach. Also the number of needed wavefront sensors and
detectors is reduced, reducing the detector read-out-noise and the
needed computing
power compared to the SO approach.\\
\begin{figure*}%[t]
 \begin{center}
  \vspace{0.3cm}
 \resizebox{2\columnwidth}{!}{
  \includegraphics[width=8cm, angle=180]{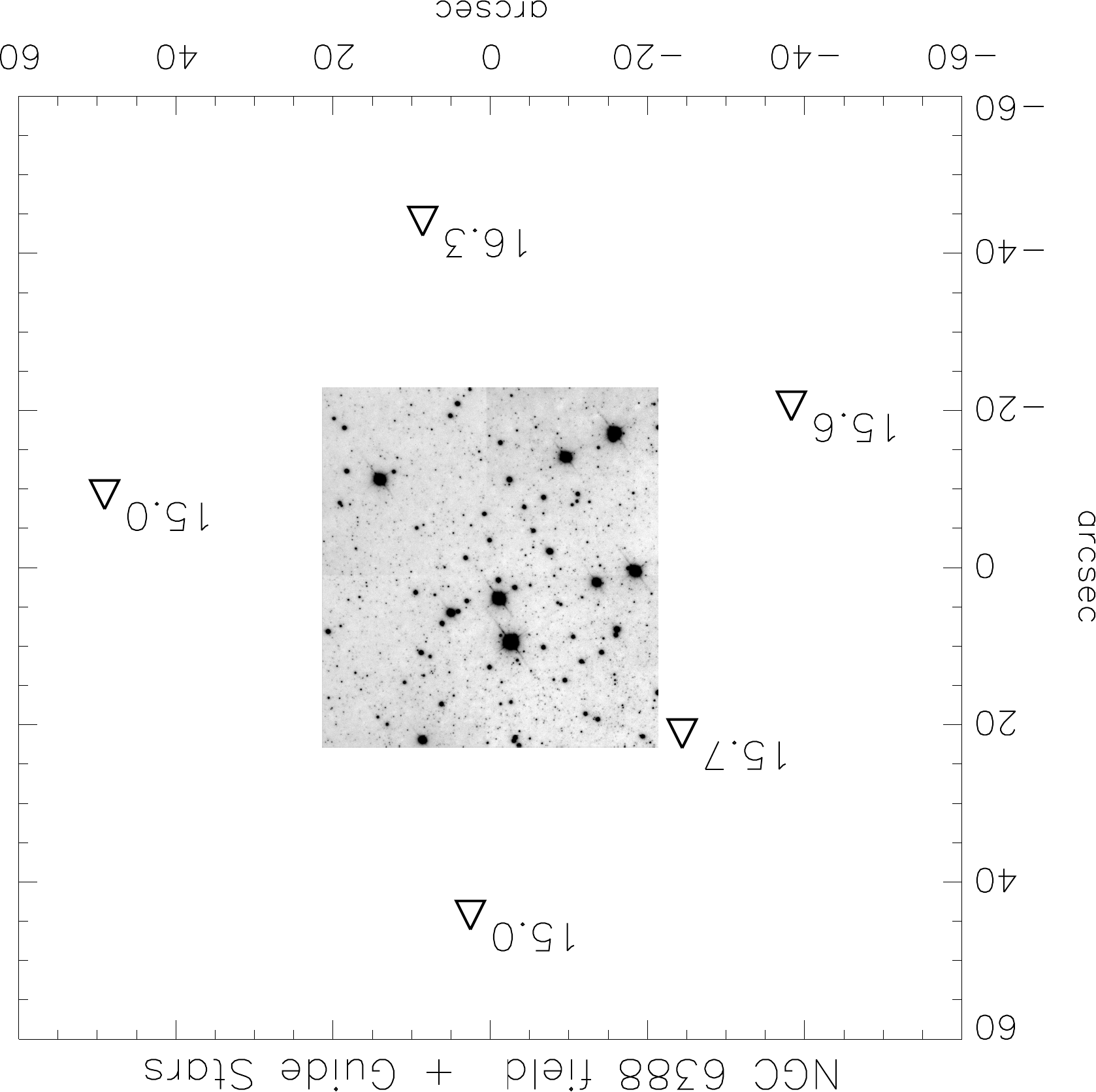} \hspace{0.5cm}
  \includegraphics[width=8cm, angle=180]{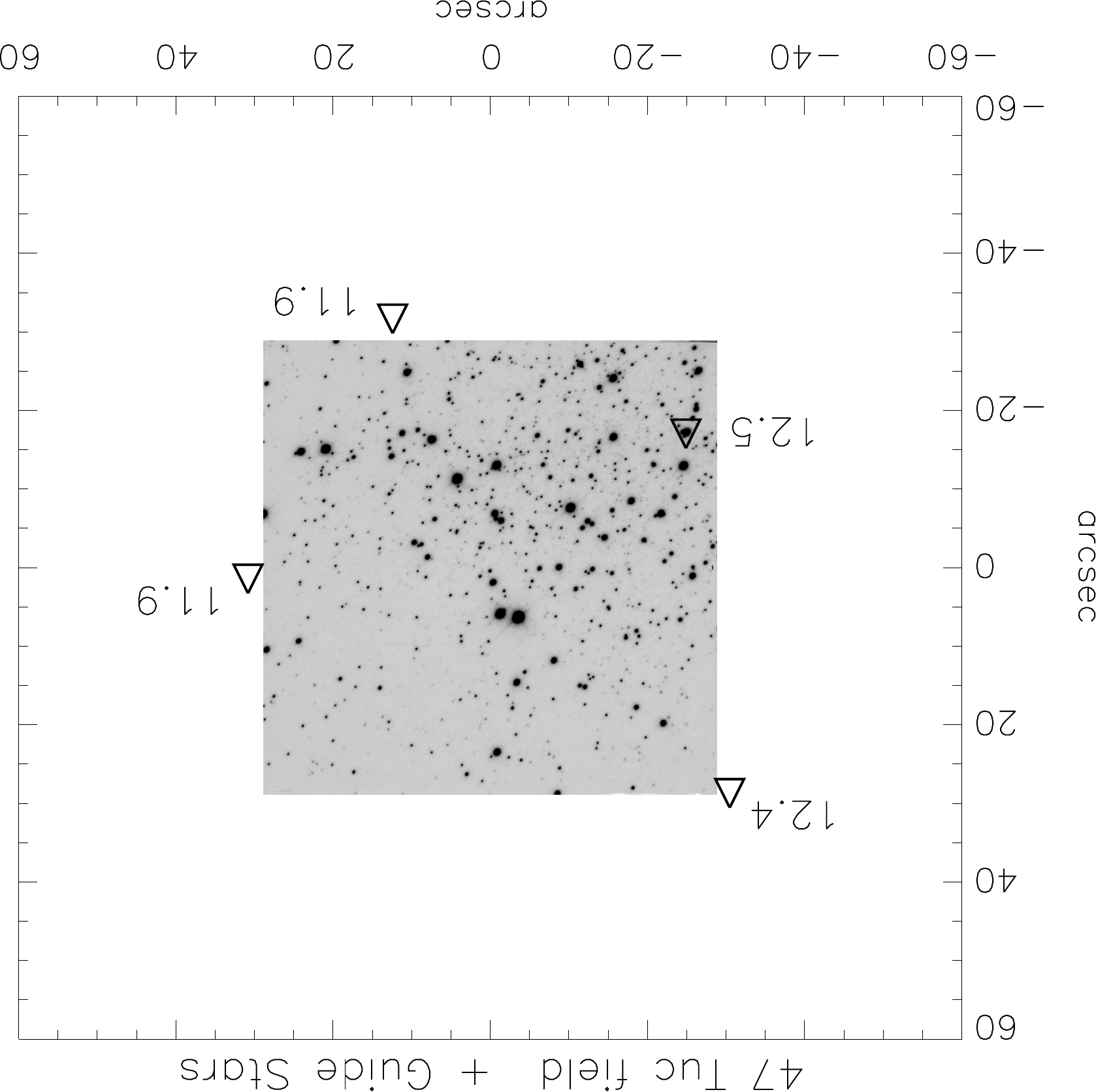}}
 \end{center}
 \caption{MAD images of the globular clusters NGC~6388 (left) and 47~Tuc (right). The triangles mark the
 positions of the AO guide stars relative to center of the observed
 FoV. The numbers close to the stars correspond to their HST F606W (visual) magnitude.}
 \label{fig:NGC6388guidestars}
\end{figure*}
\noindent High precision astrometry combined with high angular
resolution is essential to many science cases in astronomy. E.g.
observations of stars around the supermassive black hole in the
center of our own Milky Way
\citep[e.g.][]{Trippe2008,Schoedel2009} and of the central regions
of globular clusters are only possible with space-based facilities
or adaptive optics supported observations from the ground. Multi-epoch
high precision proper motion studies with the Hubble Space
Telescope (HST) made it possible to distinguish cluster members
from foreground field stars and to study the internal dynamics and
kinematics of several globular clusters and galactic starburst
clusters \citep[e.g.][]{King2001,McLaughlin47Tuc,Rochau2010}.
Another field of high precision astrometry is the detection and
characterization of extrasolar planets by measuring the
astrometric reflex-motion of the star \citep[e.g.][FGC/HST]{Benedict2002,Bean2007}, 
\citep[][NACO/VLT, in
preparation]{Meyer2010GJ}. This is an
important technique, complementary to the radial velocity method,
the most efficient detection method so far. A larger FoV enhances
the number of usable reference stars for the measurement of the
relative astrometric motion and therefore the achievable
precision significantly.\\

\noindent The Multi-Conjugated Adaptive Optics Demonstrator (MAD)
is a prototype instrument for MCAO correction and observation and was installed at
the ESO VLT UT3 at the Paranal Observatory in 2007 \citep{Marchetti2007}.
MAD was designed to study and test different MCAO systems, both in
the lab and on-sky
\citep{Hubin2002,Marchetti2003,Arcidiacono2006}. MAD 
employs adaptive optics sensing and correction in the star-oriented
and the layer-oriented mode. 
Two layers are sensed and corrected in the full MCAO mode. The
ground-layer at
the telescope's pupil and a high layer at 8.5~km altitude.\\
\noindent Future AO instruments will
use the MCAO technique, such as the Gemini MCAO System (GeMS) at
the Gemini South Observatory on Cerro Pachon, Chile. The
Fizeau-Interferometer LINC-NIRVANA for the Large Binocular
Telescope (LBT) on Mt. Graham in Arizona, will be equipped with
four layer oriented correction units, two for each telescope,
which will correct the ground layer and a high layer
\citep[e.g.][]{Farinato2008}. One of the science-cases of
LINC-NIRVANA will be the detection and characterization of
extrasolar planets.\\
\noindent All the above mentioned aspects and the uniqueness of
the very first MCAO data available encouraged us to analyse this
data in terms of astrometric precision and stability. The aim of
this study is the estimation of the achievable precision and
stability in astrometric measurements obtained with MCAO imaging.

%__________________________________________________________________

\section{Observations and data reduction}
The observations analyzed here were conducted with the multi
pyramid wavefront sensor of the MAD instrument in the LO mode
\citep{Ragazzoni1996,Ragazzoni2000,Arcidiacono2008}. This sensor
has the advantage that it can use up to 8 guide stars
simultaneously which can be relatively faint ($V < 18$) and for
which the integrated light reaches $V = 13$. A uniform
distribution of these stars is preferable but they can be
everywhere in the $2\arcmin \times 2\arcmin$ FoV. A NIR-
science camera is used for the observations, which is also used in
the SO mode: CAMCAO\,=\,CAmera for MCAO. It has a $57\arcsec
\times 57\arcsec$ field of view but can scan a circular FoV
of 2 arcmin diameter. The HgCdTe HAWAII2 IR-detector built by
Rockwell has $2048 \times 2048$ pixels with a pixel-scale of
0.028\arcsec/px, a readout-noise of 13.8 erms, a full well
capacity of 65000~ADU and a loss of linearity above 35000~ADU.
Two data sets were analyzed, one in the globular cluster NGC~6388
and the other one in the globular cluster 47~Tuc. Both data sets
are test data obtained during the first on-sky test of the LO
correction mode
with MAD at the VLT.\\
The goal of the observations was to verify and show the
capabilities of MCAO observations in LO mode. The original focus was
the photometric analysis, as especially high precision photometric
studies in crowded fields, like clusters, benefit from the large
AO-corrected FoV. Therefore note that the  observations analyzed here were not obtained
under the aspect of high precision astrometry. Nevertheless they
represent a unique data set to investigate the possibilities of
high precision astrometry with MCAO.

\subsection{MCAO - NGC~6388}
The data of the globular cluster NGC~6388 were obtained on
September $27^{th}$ 2007 using the full MCAO capability of MAD.
The observations were made in the $K_{s}$ band (central wavelength
$= 2.12~\mu m$) using 5 guide stars with $V = 15.0, 15.0, 15.6,
15.7$ and 16.3 mag\footnote{HST F606W photometry data},
corresponding to an integrated  magnitude of 13.67
\citep{Arcidiacono2008}. The guide stars are positioned around the
field of view, see Fig.~\ref{fig:NGC6388guidestars}, left. The
observed field lies at the south-eastern rim of the cluster at
RA(J2000)=17:36:22.86, DEC(J2000)=-44:45:35.53. Altogether 30
frames were obtained, the first five in GLAO mode and the last 25
in full MCAO mode. A jitter pattern of five positions was used,
repeated six times with three slightly different central points,
to scan part of the $2\arcmin \times 2\arcmin$ FoV, avoid bad pixel
incidents and for better sky estimation. The first 10
frames were obtained with a Detector Integration Time of DIT = 10
seconds and N = 24 of these DITs (=NDIT) are directly co-added
onto one frame, resulting in 240 seconds total exposure time per
frame. In the last 20 frames the number of exposures was reduced
to NDIT = 12, resulting in 120 seconds total integration time per
frame. 
In Table~\ref{tab:observations} the observations are
summarized together with performance indicators such as the FWHM
of the fitted PSF and the seeing measured by the DIMM monitor.
The same data was also analyzed for photometry by \citet{Moretti2009}.\\

\begin{table}
\caption{Summary of the observations of the clusters NGC~6388 and 47~Tuc.}
\label{tab:observations}
\begin{center}
\small{
\begin{tabular}{|c|c|c|c|c|c|c|} \hline %\vspace{0.4cm}
& \multicolumn{3}{c|}{NGC~6388} & \multicolumn{3}{c|}{47~Tuc}\\ \cline{2-7}
frame & ExpT & seeing & FWHM & ExpT & seeing & FWHM\\
& [sec] & $V$ [$\arcsec$] & $K_{s}$ [$\arcsec$] & [sec] & $V$ [$\arcsec$] & Br$\gamma$ [$\arcsec$]\\ \hline
1  & 240 & 0.43 & 0.098 & 30 & 1.09 & 0.178\\
2  & 240 & 0.41 & 0.094 & 30 & 1.15 & 0.186\\
3  & 240 & 0.49 & 0.099 & 30 & 1.13 & 0.206\\
4  & 240 & 0.55 & 0.094 & 30 & 1.08 & 0.169\\
5  & 240 & 0.51 & 0.090 & 30 & 1.09 & 0.178\\
6  & 240 & 0.41 & 0.095 & 30 & 1.08 & 0.147\\
7  & 240 & 0.38 & 0.097 & 30 & 1.15 & 0.157\\
8  & 240 & 0.37 & 0.098 & 30 & 1.17 & 0.193\\
9  & 240 & 0.39 & 0.103 & 30 & 1.17 & 0.173\\
10 & 240 & 0.40 & 0.106 & 30 & 1.15 & 0.178\\
11 & 120 & 0.45 & 0.126 & 30 & 1.14 & 0.145\\
12 & 120 & 0.43 & 0.117 & 30 & 1.15 & 0.148\\
13 & 120 & 0.45 & 0.130 & 30 & 1.11 & 0.144\\
14 & 120 & 0.50 & 0.158 & 30 & 1.15 & 0.166\\
15 & 120 & 0.51 & 0.130 & 30 & 1.14 & 0.183\\
16 & 120 & 0.49 & 0.155 & 30 & 1.15 & 0.183\\
17 & 120 & 0.48 & 0.170 & 30 & 1.19 & 0.187\\
18 & 120 & 0.49 & 0.140 & 30 & 1.13 & 0.200\\
19 & 120 & 0.45 & 0.119 & 30 & 1.11 & 0.174\\
20 & 120 & 0.41 & 0.133 & & &\\
21 & 120 & 0.42 & 0.120 & & &\\
22 & 120 & 0.44 & 0.131 & & &\\
23 & 120 & 0.54 & 0.139 & & &\\
24 & 120 & 0.56 & 0.158 & & &\\
25 & 120 & 0.43 & 0.176 & & &\\
26 & 120 & 0.50 & 0.153 & & &\\
27 & 120 & 0.46 & 0.143 & & &\\
28 & 120 & 0.48 & 0.135 & & &\\
29 & 120 & 0.47 & 0.120 & & &\\
30 & 120 & 0.47 & 0.134 & & &\\ \hline
\end{tabular}
\tablefoot{
The seeing values are measured by the DIMM seeing
monitor in the V band and the FWHM value corresponds to the one
measured in the extracted PSF, used to fit the positions of the stars.
In the case of the NGC~6388 data, the first five frames are taking
using only ground-layer correction and frames 6--30 are in full MCAO mode.
In the case of the 47~Tuc data all frames are taken using ground-layer correction.}}
\end{center}
\end{table}

\subsection{GLAO - 47~Tuc}
The observations of the globular cluster 47~Tuc were obtained on
September $22^{nd}$ 2007 using only the Ground Layer Adaptive Optics
approach \citep{Arcidiacono2008}. The center of the cluster
(RA(J2000)=00:24:05.6, DEC(J2000)=-72:04:49.4) was observed with
the narrow-band Br$\gamma$ filter (central wavelength $= 2.166 ~\mu m$) and 4
guide stars between $V = 11.9$ mag and $V = 12.5$ mag, positioned
around the field with one guide star in the south-eastern corner
of the field (Fig.~\ref{fig:NGC6388guidestars}, right side). We
have analyzed 19 frames with DIT $= 2$ sec and NDIT = 15, corresponding to
a total exposure time of 30 seconds per frame.

\begin{figure*}[!t]
 \begin{center}
  \vspace{0.4cm}
  \resizebox{1.8\columnwidth}{!}{
  \includegraphics[angle=90]{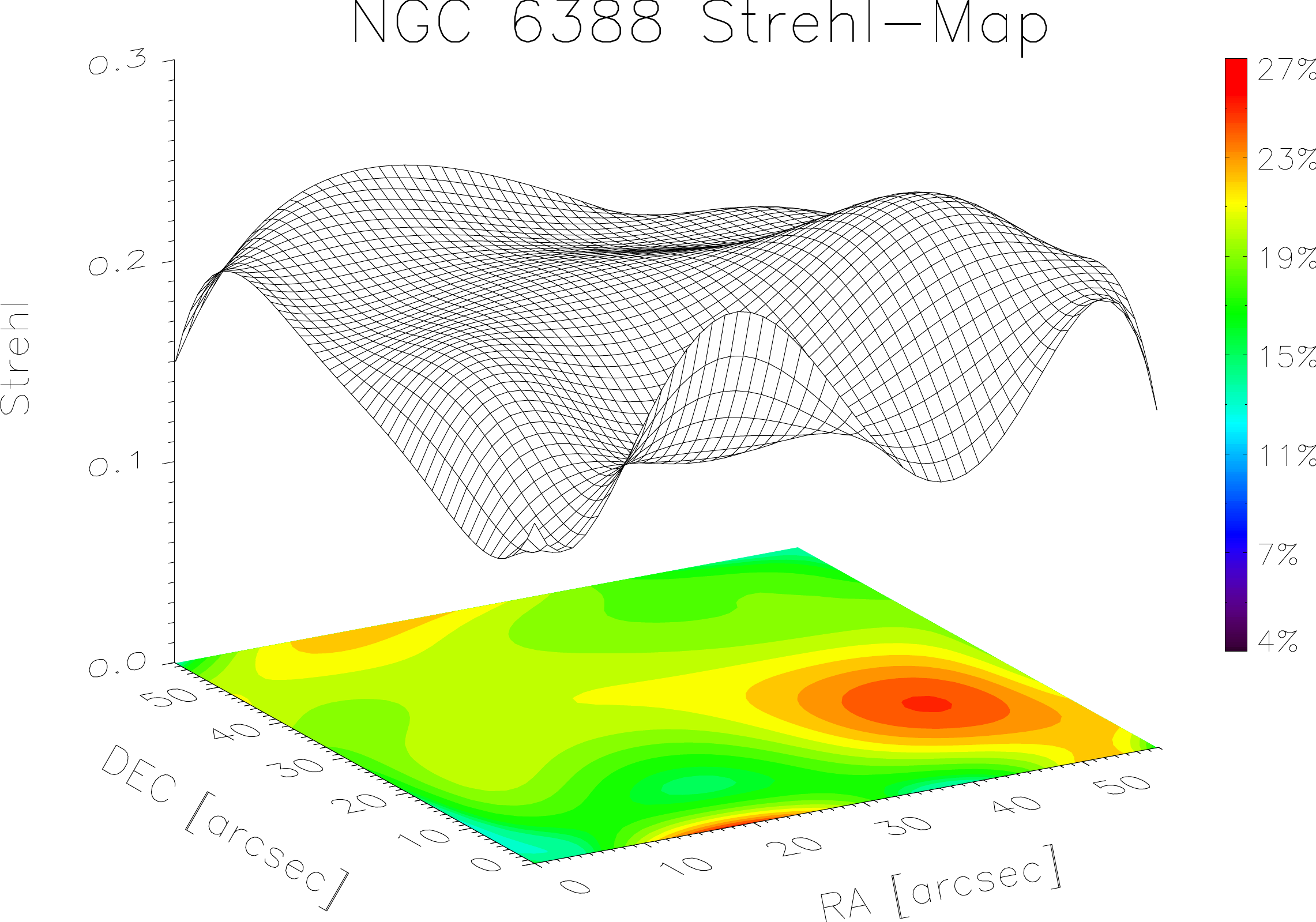} \hspace{2.5cm}
  \includegraphics[angle=90]{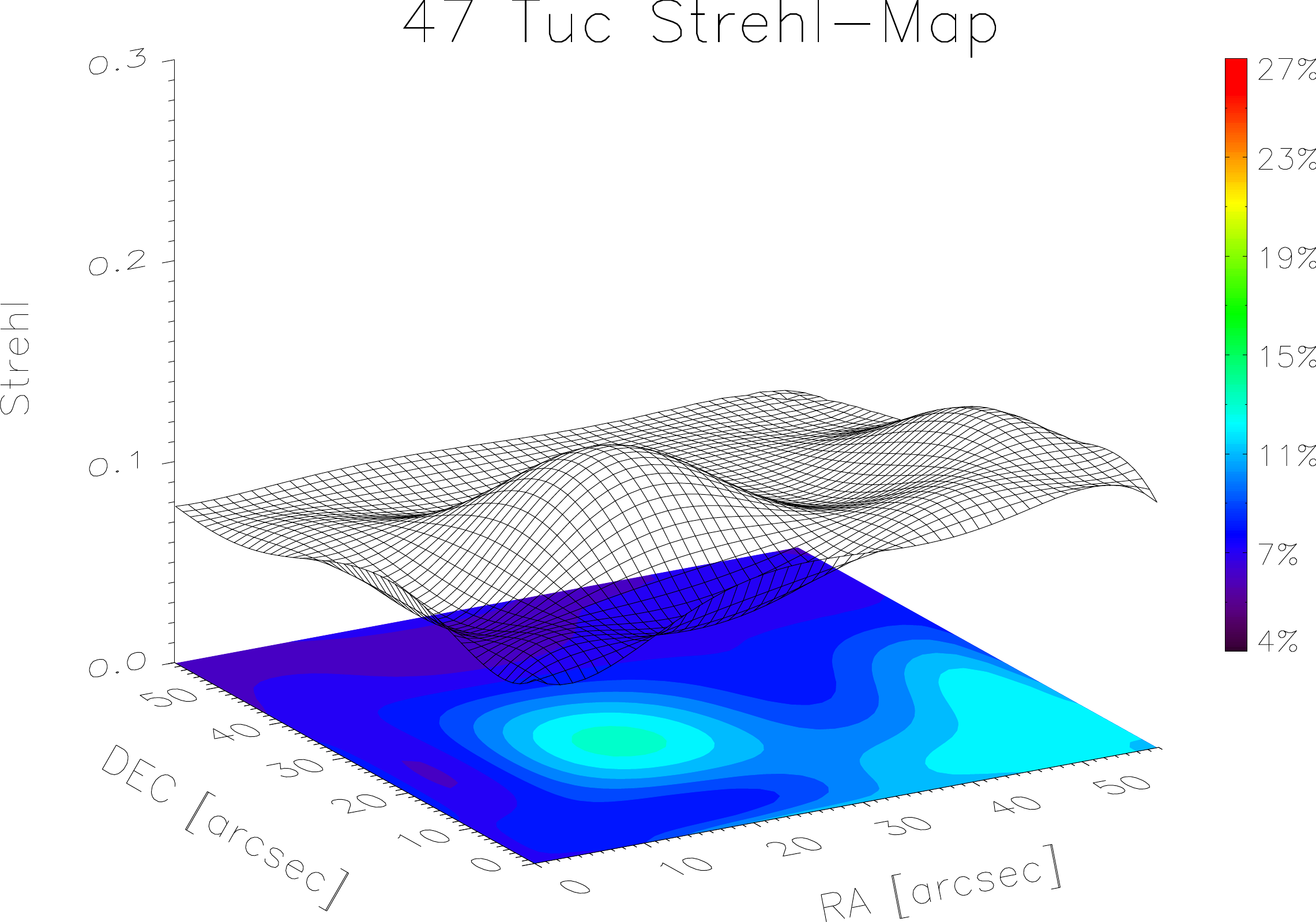}}
 \end{center}
 \caption{Example Strehl maps for one of the NGC~6388 (left) and 47~Tuc (right) data set, respectively.}
 \label{fig:Strehl}
\end{figure*}

\subsection{Data reduction}
Each science frame was flatfield corrected by the flatfield image
obtained from sky flats taken at the beginning of the night and
badpixel corrected with a badpixel mask obtained from the same
flat-field images \citep{Moretti2009}. Sky subtraction was done by
median combining all sky and science frames to one single sky
frame. This frame was then normalized to the median counts of the
science frames before subtraction. In the case of 47~Tuc no
flat-field images were taken in Br$\gamma$ (central wavelength $=
2.166~\mu m$) in that night. Consequently, we used the same
flat-field image for correction as for the NGC~6388 data observed
in $K_{s}$ (central wavelength $= 2.12~\mu m$). In the case of the
NGC~6388 data, jittering was used during the observations and we
cut all images to the common FoV after the data reduction. This
left a slightly smaller field of the size of $1517~\rm~px \times
1623~\rm~px$ ($42.5\arcsec \times 45.4\arcsec$). Only stars which
were detected in
 all frames were used in the following astrometric analysis.

\section{Strehl Maps}
\label{sec:Strehl} As a check of the AO performance we generated
Strehl-maps by calculating the Strehl-ratio for each detected
star. After interpolating values for areas where no stars were
found, a smooth surface was fitted to the data, leading to a two
dimensional Strehl-map for each frame. In Fig.~\ref{fig:Strehl}
one example of these maps is shown for each data set.

\noindent The Strehl is fairly even over the field of view, mean values are
between 10\% - 23\% in the full MCAO case and between 9\% - 14\%
in the GLAO case, with a small drop-off to the edges of the field.
This shows how uniformly the layer oriented MCAO approach corrects
wavefront distortions. The drop-off to the edges of the FoV can
partly be explained by the MCAO and the atmospheric tomography
approach. The light coming from the different directions of the
guide stars is optically co-added and a correction is computed
based on this light distribution. But the footprints of the
columns above the telescope in the direction of the guide stars
overlap more in the middle of the field than at the edges in the
higher layer. If the control software is not optimized to correct
over the whole FoV very evenly, the middle of the field will be
corrected better. As the data analyzed here is the first data of
MCAO in layer oriented mode, we are not surprised to see such an
effect. A performance evaluation of these data can be found in
\cite{Arcidiacono2008}.
\noindent In the case of 47~Tuc the Strehl
is smaller than in the case of the NGC~6388 data. GLAO works as
a seeing reduction and Strehl ratios of a few percent are expected.
The performance of the ground layer correction was therefore even
better than expected, which was most probably due to the fact that
during the ground layer observations the turbulence was particularly
concentrated in this layer. MCAO should retrieve larger and more uniform
Strehl ratios of the order of 20\%-30\% and diffraction limited FWHM
values. Therefore the MCAO corrections did not yet fully reach the
expectations. Nevertheless, an even Strehl ratio of
$\sim 10~\%$ or  more over a $1\arcmin \times 1\arcmin$ FoV is already an
enhancement compared to the seeing limited  and the single guide
star case, where the Strehl is varying with the separation to the
guide star \cite{Roddier1999,Cresci2005}.

\section{Astrometric measurements}
\subsection{Position measurements}
To measure the positions of the stars in the single images of both
clusters we used the program \textit{StarFinder}
\citep{Diolaiti2000,Diolaiti20002}, which is an IDL based code for
PSF fitting astrometry and photometry in AO images of stellar
fields. The following description accounts for both data sets.\\
We extracted the Point Spread Functions (PSF) for fitting the
stars directly from the images, by using in each frame the same 30
stars to create the PSF by averaging these stars after deleting
close secondary sources. The selected stars are evenly distributed
over the FoV and are composed of brighter and fainter ones, with magnitude
ranges of $K = 11 - 15.3$ in the NGC~6388 data and $K = 6.3 - 12.2$ in
the 47~Tuc data. We
assumed here, that the PSF does not vary strongly across the FoV.
Analysis and tests we performed on the distributions of the
eccentricity and orientation of the PSFs in the full FoV did not
show a prominent variation over the field. The eccentricities and
the orientations of the PSFs were analyzed by fitting a
2-dimensional Moffat function to the individual PSFs with the IDL
based non-linear least square fitting package mpfit2dpeak,
provided by Craig Markwardt \citep{Markwardt2009}. Results from e.g.
\citet{Schoedel2010} and \citet{Fritz2010} show that the PSF
variation due to anisoplanatism can add an error to the position
measurement of up to 0.1 pixel. These numbers were calculated
for a classical AO correction with the S27 camera of the VLT/NACO
instrument which has a similar image scale as the MAD detector
(27.15 mas/px (NACO) vs. 28 mas/px (MAD)), but uses only one AO reference star,
located somewhere in the FoV. The data analyzed here are the first ones
obtained with multi-conjugated AO correction in the layer-oriented
approach. These two circumstances lead to a more uniform PSF over
the full FoV, as can also be seen in the even Strehl distribution
(see \S~\ref{sec:Strehl}). Another confirmation that this
assumption is acceptable can be seen in the case of the 47~Tuc
data set. One of the used guide stars lies within the FoV
(south-eastern corner, see Fig.~\ref{fig:NGC6388guidestars} right
side). Inspection of the eccentricity and orientation of the PSFs
shows that the guide star does not differ in shape and orientation
from the other stars. A behaviour, such as the change of the PSF
dependent on the separation of the stars from the guide star, as in
the classic AO
correction, cannot be seen.\\
After deleting false detections we matched the starlists to find
the stars common to all frames. This left $\sim$ 130 stars for the
NGC~6388 field and $\sim$ 280 stars for the 47~Tuc field.\\

\subsection{Distortion correction}
\label{sec:Distortion} To investigate the stability of the MCAO
and GLAO performance in terms of astrometric precision over time,
we first corrected for distortions of the field. During the
observations, malfunctioning of the de-rotator occurred due to a
software problem, leading to a bigger rotational error in several
frames. If the AO correction is very stable over time, the
relative positions of the stars should be the same after
correcting for effects such as the de-rotator problem \citep{Arcidiacono2010}. A misposition of the
reference star on the tip of the pyramid-WFS exceeding a few $\lambda/D$
(where $\lambda$ is the wavelength and $D$ the telescope diameter) with respect to
the theoretical (unrotated) positions, affects also the closed loop
performance, generating a correction under-performance. We set up a master-coordinate-frame to which
the single frame coordinates are later mapped. In order to create
this coordinate frame, we used the best frame, chosen
according to the highest mean Strehl ratio in the images, as a
first reference frame and mapped all the stellar positions from each
individual frame onto this reference frame by calculating the
shift and scale in $x$- and $y$-direction and the rotation between
these frames. The
MIDAS\footnote{http://www.eso.org/sci/data-processing/software/esomidas/}
data reduction software and simple affine transformations were
used for the transformations. We did not apply any interpolation
directly to the images, but instead worked with the measured
coordinates. 
After correcting for the derived rotation for each frame, as well as
for the shift and scale in $x$ and $y$ of each stellar position, a
master-coordinate-frame was created by averaging the position of
each star over all frames. The so derived coordinate frame with
averaged positions was then used as the master-coordinate-frame
for the coming
analysis.\\
\begin{figure*}
 \begin{center}
  \vspace{0.3cm}
   \resizebox{1.8\columnwidth}{!}{
   \includegraphics[width=7cm]{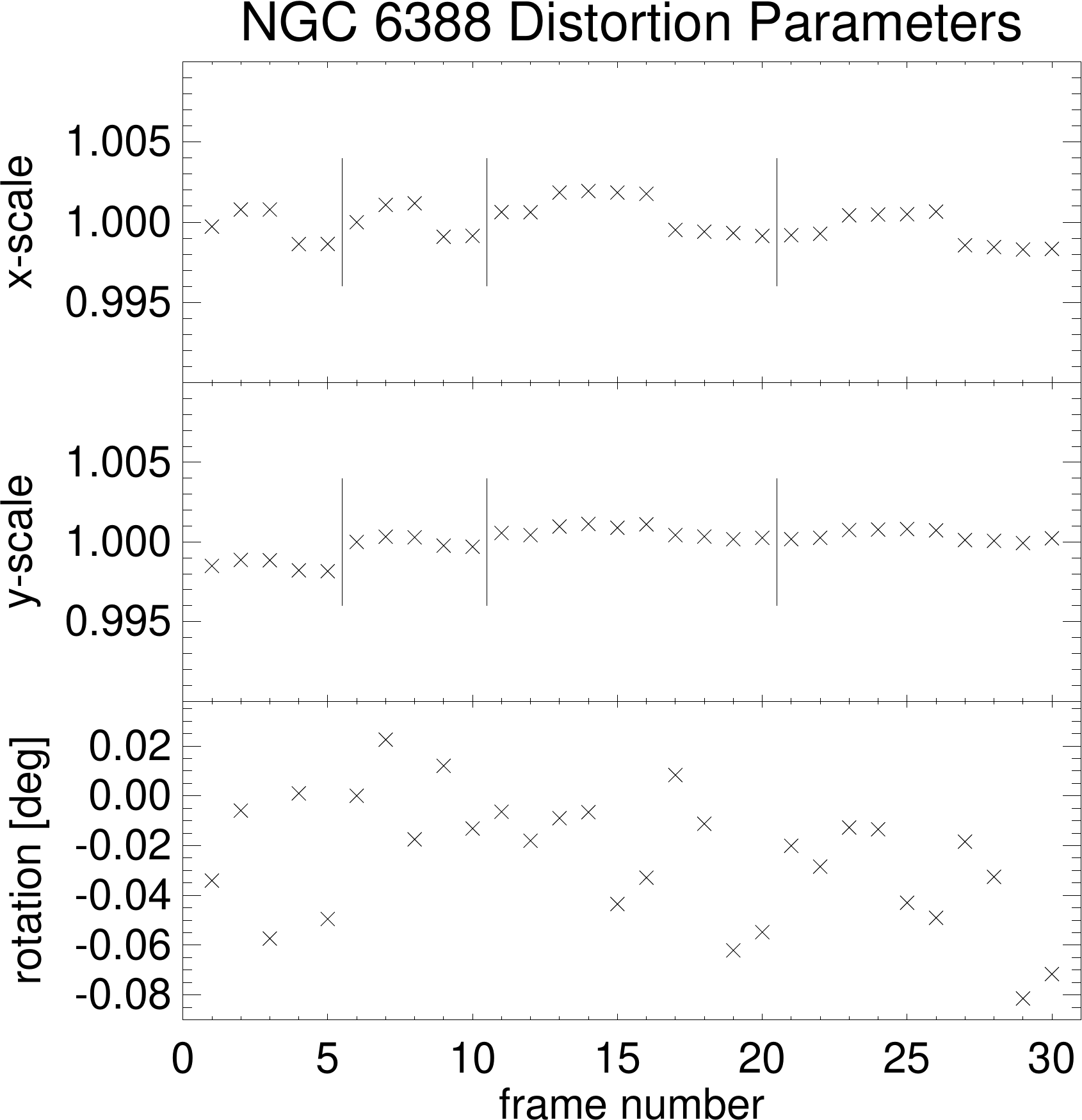} \hspace{1cm}
   \includegraphics[width=7cm]{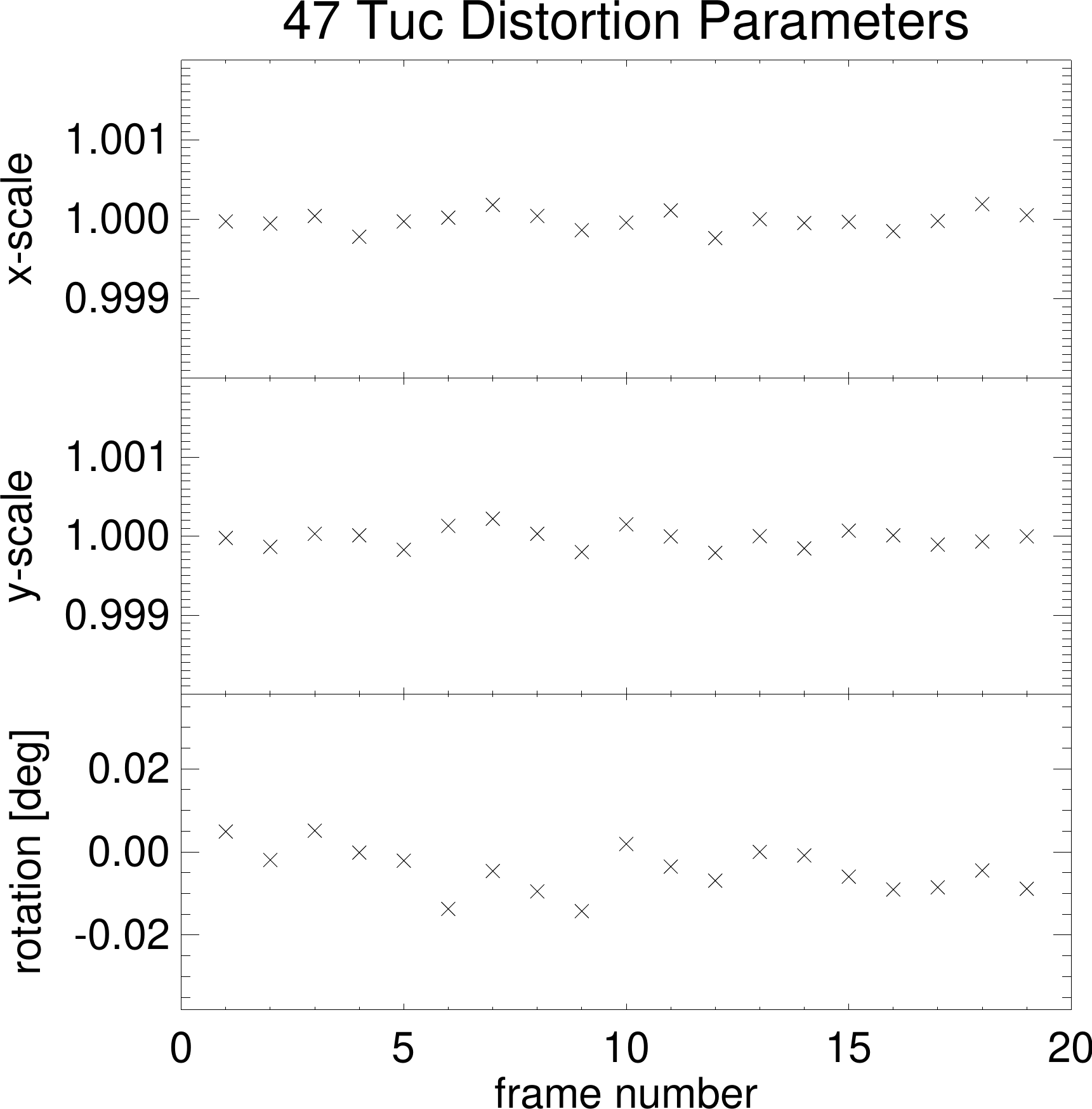}}
 \end{center}
\caption{Applied distortion parameters for the basic distortion 
correction over frame number for the
NGC~6388 (left) and 47~Tuc data (right). The panels show from top
to bottom the calculated distortion parameters for $x$ scale, $y$
scale and rotation for each frame. While the rotation parameter is
random, the scale parameter of the NGC~6388 data shows a pattern
which is not visible in the 47~Tuc data. The vertical
lines indicate after which frame the jitter movement of five
positions (five or ten frames) was repeated and highlight the
introduced pattern.} 
\label{fig:distortions}
\end{figure*}

\noindent In the following,
regarding the necessary distortion correction of each single frame, we 
analyzed the data using two approaches.\\
\noindent In a first attempt we corrected only basic distortions, including
shift, scale and rotation. To furthermore explore the full capacity of
astrometry with MCAO, we performed a second distortion correction which included
higher order terms.

\subsubsection{Basic distortion correction}
\label{sec:basic_dist}
\noindent Once we created the reference frame for each data set, all
coordinates from each single frame were then mapped
to this master-frame, leading to a better calculation of the transformation
parameters for the individual frames. One might
think that one can achieve even better transformations between the
frames by applying this method iteratively, creating once more a
master-coordinate-frame. If the distortions in the images, those
left over from the AO or systematic ones, were homogeneous over
the FoV, the transformations should not change or enhance the
positions of the master-coordinate-frame a lot. But if the
distortions are not homogeneous, but depend, for example, on the
camera position in the FoV, one would introduce warpings in the
master-coordinate-frame which one cannot map with a simple
combination of shifting, scaling and rotation anymore. We
therefore stopped after one
iteration.\\
\noindent We then calculated the residual separations between the positions of
the stars in the master-coordinate-frame to the positions in
the individual frames, calculated with the obtained transformation
parameters and analyzed them as a measure of astrometric precision.\\

\subsubsection{Separation Measurements}
\label{subsec:distMeas} To evaluate the astrometric precision and
stability of MCAO data, we measured the relative separations
between various pairs of stars all over the FoV before and after
applying the calculated distortion corrections. For this we
derived a time sequence of the separation over all frames. If only
a steady distortion were present in the single frames, then the
separations should be stable over time or only scatter within a
certain range given by the accuracy of the determination of the
position of the stars, which is 0.33~mas for the faintest star
used in this analysis. If differential distortions between the
single frames are present, but these distortions are random, the
scatter of the separations is expected to increase, depending on the
strength/amplitude of the differential distortions. A not perfectly
corrected defocus, for example, would change the absolute
separation between two stars, but, to first order, not the
relative one measured in the individual frames, if this defocus is
stable over time. An uncorrected rotation between the frames would
change the separation of two stars in the x
and y direction, but not their separation, $r = \sqrt{\Delta x^2 + \Delta y^2}$.\\

\noindent Performing this test for several star pairs with short and
large separations and with different position angles between the
stars before any distortion correction, showed in the case of the
NGC~6388 data a recurring pattern
in the separation in $x, y, r$, which is not observable in the
47~Tuc data. Fig.~\ref{fig:separation} shows the separation in $x,
y$ and $r$ over the frame number for five representative pairs of
stars in the NGC~6388 data. Looking at the pattern, which repeats
after five frames for the first 10 frames and after 10 frames for
the following frames (where always two images were taken at the
same jitter position  before moving to the next position), one
finds that this change in separation seems to be correlated with
the jitter movement during the observations which also has a five
points pattern with an additional change of the center position. In
the case of the MAD instrument the camera itself is moved in the
focal plane to execute the jitter pattern. This can lead to
vignetting effects for larger jitter offsets and distortions seem
to be introduced, dependent on the position of the camera in the
field of view. It is unlikely that this pattern is due to the
problems with the de-rotator, because of the uniform repetition of
the pattern. Also this pattern is not seen in the 47~Tuc data,
which was obtained without jitter movements, but experienced the
same
de-rotator problems.\\
\noindent We performed the same measurements of the same star
pairs after applying the calculated distortion correction for
shift, scale and rotation. The
strong pattern is gone, leaving a more random variation of the
separation. Also the calculated standard deviation is much
smaller, ranging from a factor of $\sim 3$ up to a factor of $\sim
19$ times smaller. Comparing the single standard deviations shows
a smaller scatter among their values than before the distortion
correction. All this leads to the conclusion, that the calculated
and applied distortions remove a large amount of the separation
scatter, but not all of it. The remaining scatter of the
separations between the stars in the single frames still ranges
from $\sim 1.2 - 2.8$ mas, well above the scatter expected from
photon statistics, pointing to uncorrected higher-order
distortions.\\

\subsubsection{Basic distortion parameters}
\label{sec:basic_dist_param}
The calculated distortion parameters from the basic 
distortion correction for $x$-scale,  $y$-scale and
rotation over the frame number, which can be seen as a
time-series, are plotted in Fig.~\ref{fig:distortions} for both
data sets. Whereas the parameter for the rotation correction looks
random, but with a fairly large scatter indicating the de-rotator
problem, the correction parameters for the scale in $x$ and $y$
show a pattern in the case of the NGC~6388 data set (left). This
pattern repeats after five (10) frames, as does the pattern for
the separation measurement. As these are the applied correction
parameters, they nicely show the existence of the pattern and our
ability to correct for this induced scale variation due to the jitter
movement. In the 47~Tuc data there is also some scatter, which can be
expected, but no repeating pattern can be seen. Additionally,
the values are smaller in the case of these ground layer corrected
data which were obtained without jitter (note the different scaling
of the two plots).

\subsection{Higher order distortion correction}
We additionally performed a comparison of the individual frame stellar 
positions with the reference positions by applying a polynomial fit 
including also higher order transformation terms. To derive the transformation 
coefficients, we used the IDL routine \textit{POLYWARP}, which is able
to fit a polynomial function of several orders, using a least squares
algorithm \citep[see e.g. also][]{Schoedel2009}. The polynomial functions used are:
\begin{align}
 X_{i} = \sum_{i,j} Kx_{i,j}X_{0}^j Y_{0}^i \\
 Y_{i} = \sum_{i,j} Ky_{i,j}X_{0}^j Y_{0}^i
\label{equ:xi}
\end{align}
where $X_i, Y_i$ are the reference stellar positions and $X_0, Y_0$ the 
stellar positions in the individual frames. $Kx_{i,j}, Ky_{i,j}$ are the 
coefficients to be calculated. 
After performing the fit with orders from $i,j \leq 1-10$, we decided to
perform 
the final transformation with an order of $i,j = 4$.\\
A fit of order four gives an enhancement of 6-10\% (NGC~6388) and 14-19\%
(47~Tuc)
of the remaining mean residuals compared to the fit of order 3. Fitting even 
higher orders is not necessary, as no significant enhancement of the 
residuals can be seen.\\
\noindent After transformation of the stellar positions in the individual frames
to the common reference frame, the residual separations were calculated as 
in the case of the simpler transformations, see \S~\ref{sec:basic_dist_param}.
\begin{figure*}
 \begin{center}
  \vspace{0.3cm}
  \resizebox{2\columnwidth}{!}{
  \includegraphics{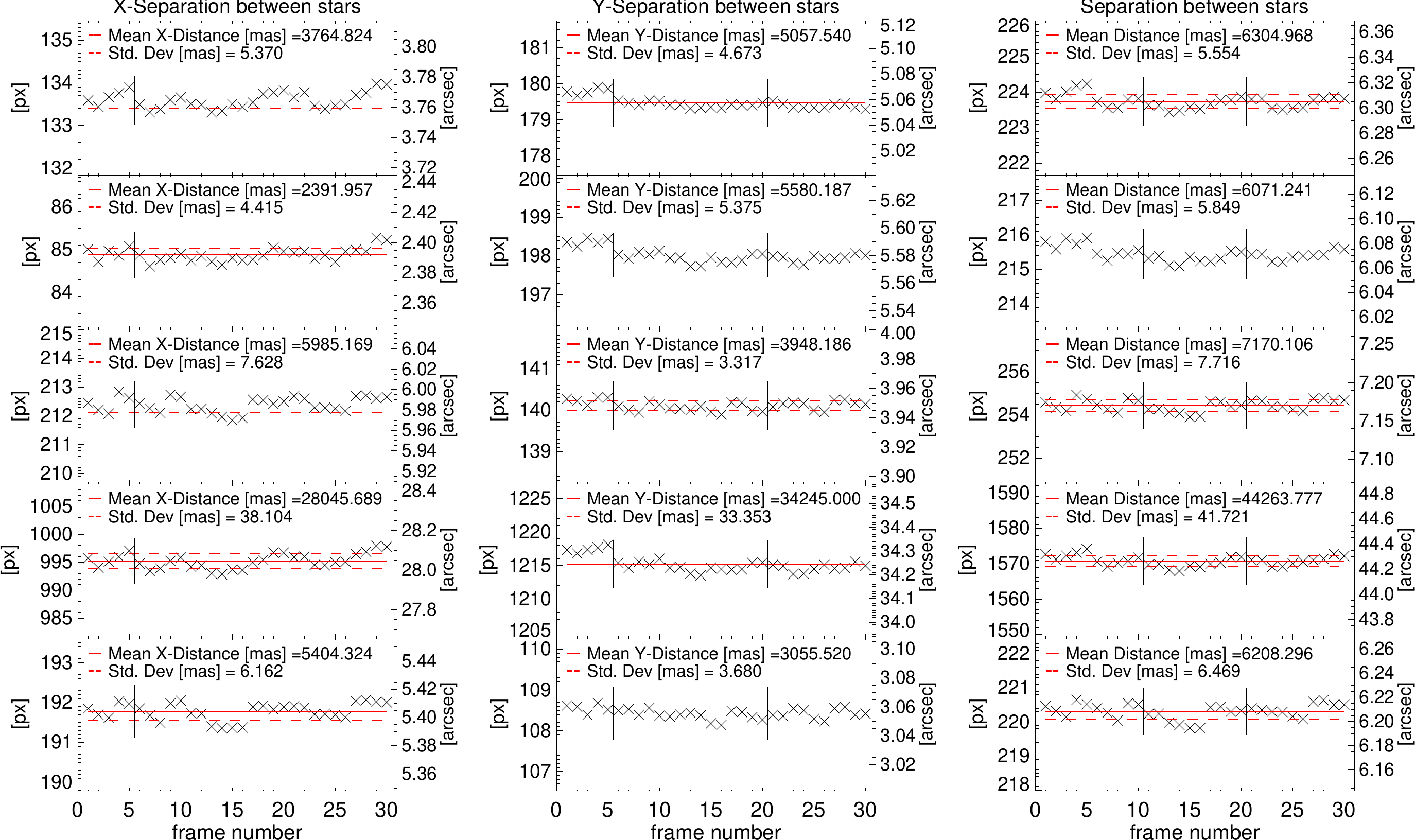}}
 \end{center}
 \caption{Separation between pairs of stars in the NGC~6388 data
 set plotted against frame number before any distortion correction. The
left panel
 shows the separation in x-direction, the middle panel in
 y-direction and the right panel the full separation
 $r = \sqrt{\Delta x^2 + \Delta y2}$. The left $y$-axes
are given in \emph{pixel} while the right ones give the measured
distances in seconds of arc. The small straight
 lines mark the frames after which a new five points
 sequence of jitter movements was started. The $x$-axes can
 also be seen as a time sequence as the individual frames were
 obtained subsequently, the first ten with an exposure time of
240~s and the last 20 with an exposure time of 120~s.}
\label{fig:separation}
\end{figure*}

\section{Results}
The basic distortion corrections are sufficient for high accuracy
photometry \citep{Moretti2009}, but to achieve the highest astrometric precision, a distortion
correction including higher orders is necessary. We therefore performed a higher
order distortion correction and show the compelling results in the
following paragraph.

\subsection{Residual Mapping}
\label{sec:residuals} After calculating the residuals for each
frame with respect to the master-coordinate-frame, we analyzed the
distribution of these residuals over the field of view.\\
\noindent We analyzed contour plots of
the residuals by fitting a minimum curvature surface to the data
of each frame to look at the spatial distribution
of the residuals after the distortion correction 
The main goal of this test was to check for any strong spatial
variation of the residuals over the FoV. No strong spatial
variation can be seen, such as for example a strong gradient in
one direction.
Additionally we analyzed arrow diagrams showing not only the
strength, but also the direction of the residuals for each star
used to calculate the transformation. We found the orientation of
the arrows to be random. \\

\begin{figure}[!ht]
 \begin{center}
  \includegraphics[width=7.4cm]{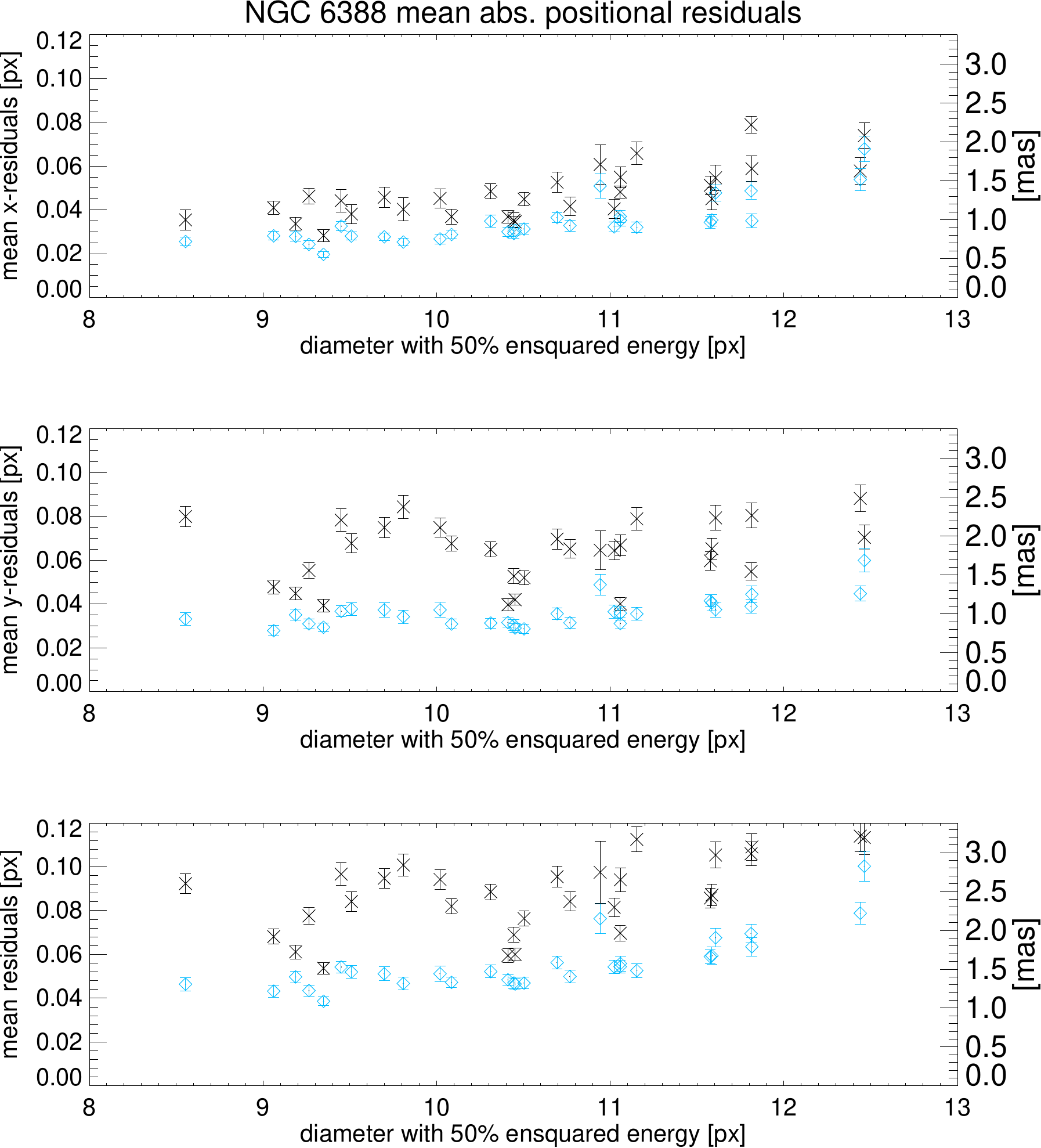} \hspace{3.5cm}
  \includegraphics[width=7.4cm]{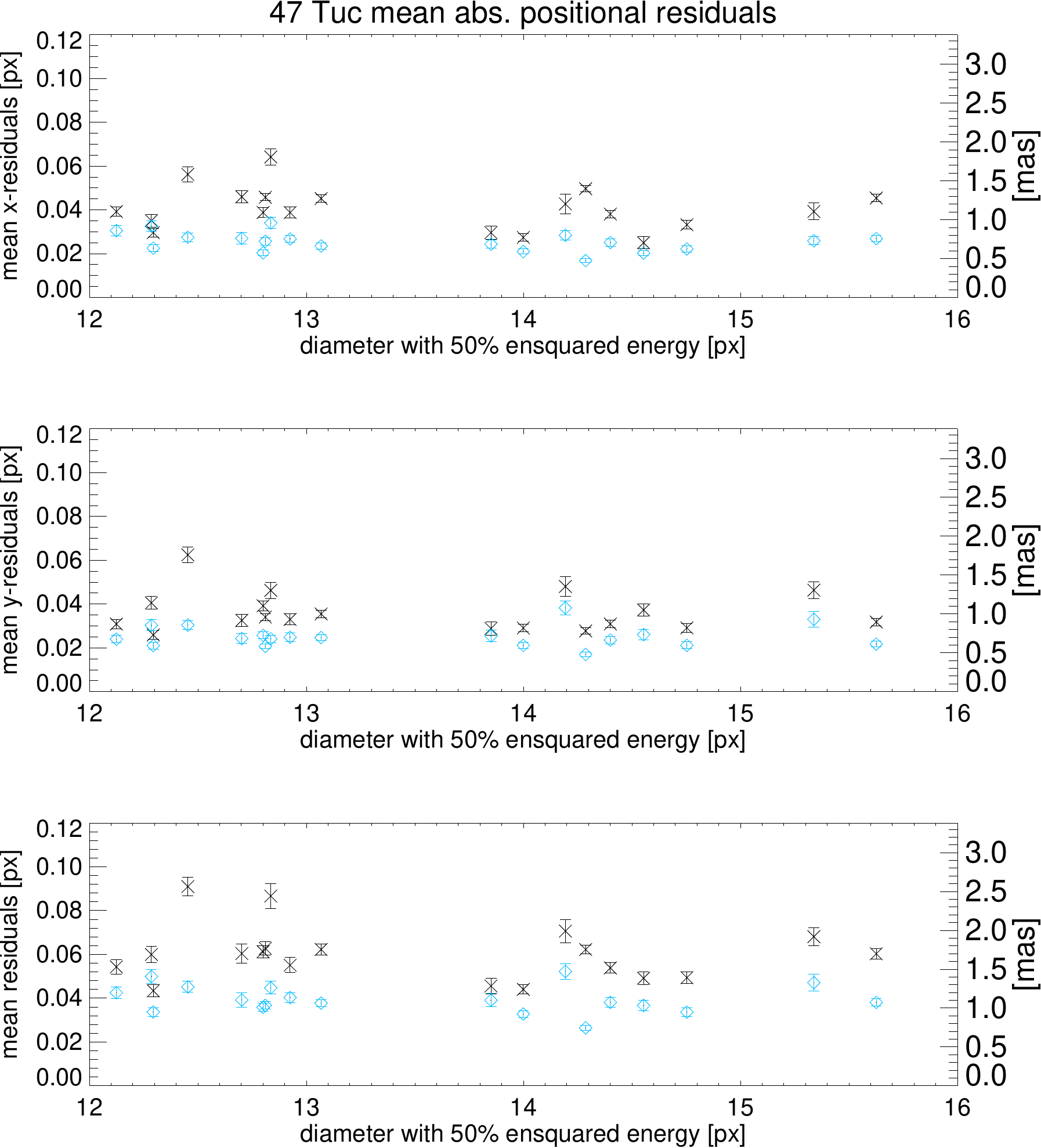}
 \end{center}
 \caption{Mean of the absolute values of the positional residuals
 over the diameter of 50\% ensquared energy. In the upper panels for the NGC~6388
 data set and in the lower panels for the 47~Tuc data set. The plot shows
 from top to bottom the mean of the absolute values of the residuals to the masterframe in the
 $x$- and $y$-direction and the separation $r = \sqrt{x^2 + y^2}$
 after a $4^{th}$ order polynomial correction
 (diamonds $\lozenge$) and after the correction of $x$ and $y$-shift, $x$ and
$y$-scale and rotation (crosses $\times$). The overplotted error
 bars correspond to the error of the mean value $\sigma /\sqrt{n}$,
 with $n$ equal to the number of stars used to calculate the mean value and
 $\sigma$ being the standard deviation. The left $y$-axis shows the
 residuals in units of pixel and the right one in units of mas.}
 \label{fig:resid_frame}
\end{figure}

\noindent Finally, we calculated the mean residuals over the full
FoV for both data sets separately for the $x, y$ and $r$ $\left ( r =
\sqrt{\Delta x^2 + \Delta y^2}\right )$ direction for each frame. The mean
values are
very close to zero ($\sim 10^{-5} - 10^{-6}$~pixel), supporting the
results from the arrow plots of random orientation, but the mean
of the \textit{absolute} values of the residuals provides a better
indicator for the variability present in the data. In
Fig.~\ref{fig:resid_frame} the mean of the absolute values of the
residuals over the full FoV in the $x$ and $y$ directions and in
the separation $r$ are plotted over the diameter of 50\% ensquared
energy of the corresponding extracted PSF of each frame and each
data set. 
In the case of an image where the flux is given as flux
per pixel, like any detector image, the ensquared energy is
defined as the flux of a PSF within a certain quadratic box with
the size of $n \times n$ pixel divided by the total flux. The
smaller the side length (=diameter) of this box, containing 50\%
of the total energy, the better the AO correction, moving flux
from the wings into the core of the PSF.
No correlation of the size of the residuals with the ensquared
energy and therefore the performance of the AO system can be seen
in the 47~Tuc data set, but there is a small correlation 
in the NGC~6388 data. What is visible, is that the
absolute values of the residuals and their scatter are larger in
the case of the NGC~6388 data set compared to the 47~Tuc data set,
even though the initial observing conditions were better and the
measured FWHM values and diameters of 50\% ensquared energy are
smaller. Compared to a only basic correction of the image
distortions (crosses $\times$), the residuals after a higher order
correction (diamonds $\lozenge$) 
are noticeable smaller and do not scatter as much as in the case of the basic
distortion correction, showing the superiority of the higher order
correction.\\
\noindent The values after the higher order correction give a first impression of how precise the
astrometry is in these MAD data. The mean absolute residuals are
between 0.020~px and 0.068~px (0.55 - 1.90~mas) in the $x$-direction
and between 0.028~px and 0.060~px (0.78 - 1.68~mas) in the
$y$-direction in the MCAO corrected NGC~6388 data set. In the case
for the ground layer corrected 47~Tuc data set, the absolute
values of the residuals are between 0.017 - 0.034~px (0.47 -
0.95~mas) and 0.017 - 0.038~px (0.47 - 1.07~mas) in the $x$ and $y$
direction, respectively.\\ 
\noindent With photon statistics alone, the
positions should 
have
a smaller range of variation. Taking the positional accuracy
calculated from photon statistics for the faintest stars used in
this set, the residuals should be within 0.012~px (0.33~mas) and
0.011~px (0.32~mas) in $x$ and $y$ direction, respectively, in the
NGC~6388 case and 0.005~px (0.14 mas) for both, $x$ and $y$, in
the 47~Tuc  case. The accuracies $\Delta x, \Delta y$ from
photon statistics were thereby calculated by: $\Delta x/y =
\frac{FWHM_{x/y}}{\sqrt{n}}$, where $FWHM_{x/y}$ is the Full Width
at Half Maximum of the fitted PSF in $x$ and $y$, respectively and
$n$ the number of photons of the fitted star.\\ 
\noindent The basic distortion correction, where we only accounted
for shift, scale and rotation, 
 shows a residual positional scatter that cannot be explained
by simple statistical uncertainties. It rather shows that even
after a basic distortion correction, there is remaining positional
scatter, which seems to have its origin in higher order
distortions present in the images, as it seems largely independent
from the size of the PSF.\\ 
\noindent The residuals
after the 
higher order corrections show a significant enhancement in the precision,
even though the values are still larger than the ones taking only
photon statistics into account. But one has to remember that the latter
values only show the lower limit of
the
possibly reachable positional accuracy. In reality the uncertainties 
  will likely be larger, possibly due to locally by the AO correction 
  induced distortions, which cannot be well described by polynomials 
  ans/or errors from the PSF estimation and fit.\\ 
  
\noindent Additionally, the residuals and scatter
are larger in the case where the camera was jittering to scan a
bigger field of view. This jitter movement introduced distortions,
which were already visible in the separation measurements and in
the distortion correction parameters calculated for scale and
rotation, see \S~\ref{sec:basic_dist_param}. But also the AO correction can
introduce 
distortions, as it dynamically adapts to atmospheric turbulence
changes. With only these two data sets available in the LO
correction mode, which are suitable for this analysis, it is not
possible at this time to separate the different error sources.
Although in the case of the 47~Tuc data where no jitter movement
was performed, the residuals can be interpreted as effects of
correcting only the ground layer and the AO correction itself.
\begin{figure*}
 \begin{center}
\vspace{0.3cm}
  \includegraphics[height=6cm]{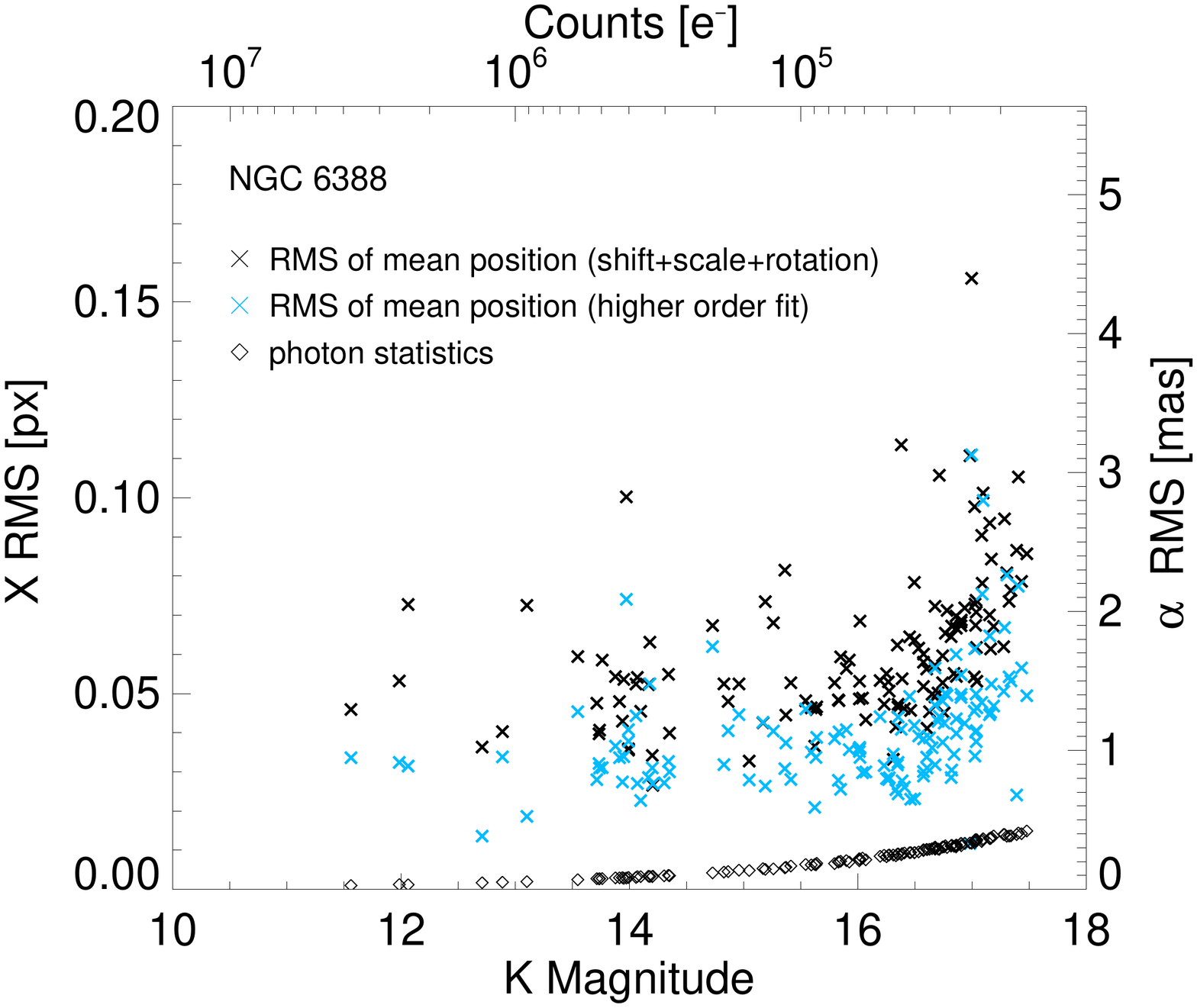}% \hspace{1.5cm}
  \includegraphics[height=6cm]{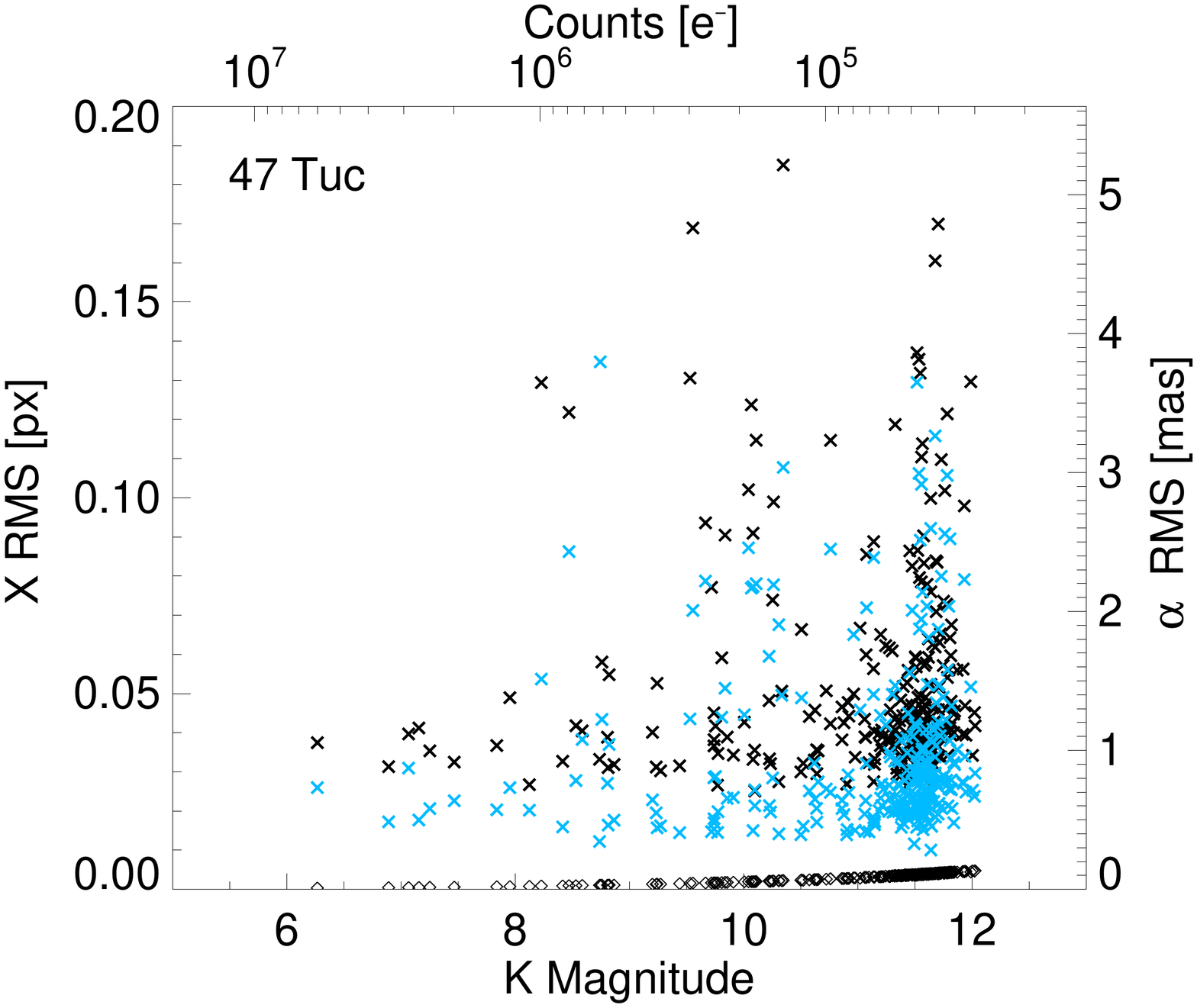} %\\ \vspace{1.5cm}
  \includegraphics[height=6cm]{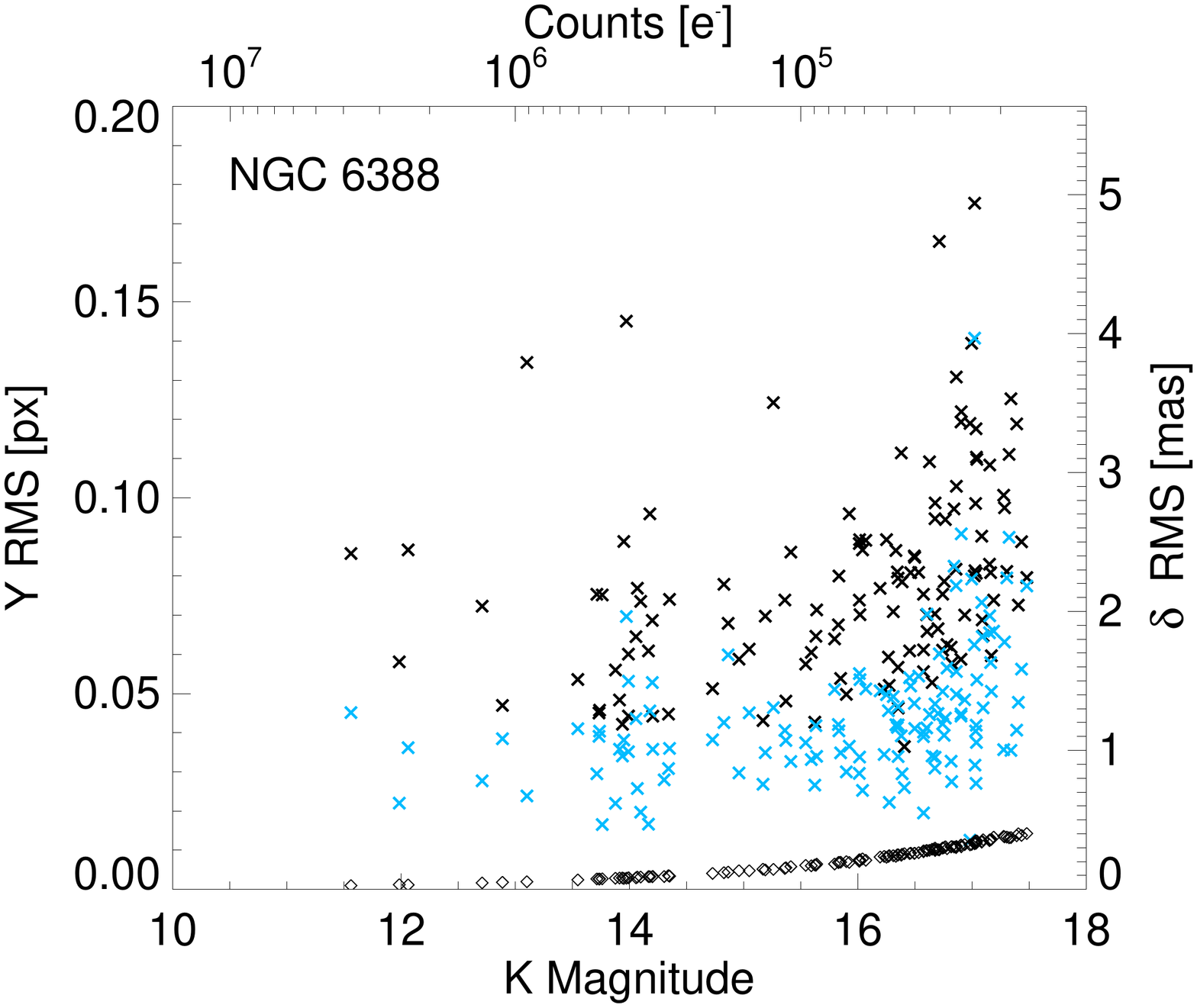} %\hspace{1.5cm}
  \includegraphics[height=6cm]{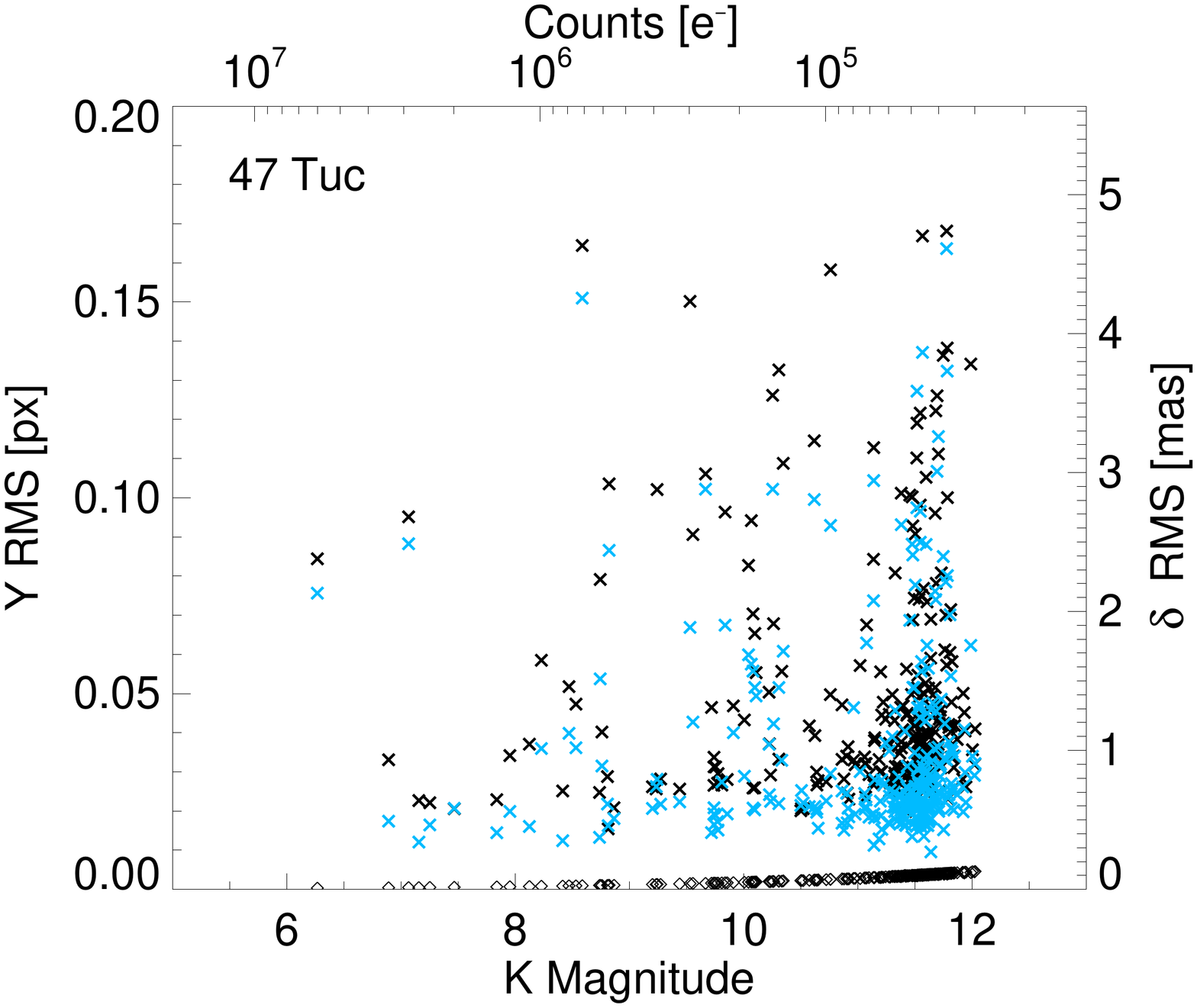}
 \end{center}
\caption{MAD positional RMS ($\times$), calculated over all frames as function
of the corresponding 2MASS $K$-magnitude for both, the higher order correction (blue)
and the basic distortion correction (black). The left panels show the
data of the NGC~6388 cluster which was observed with the full MCAO
mode for the $x$- and $y$-direction and the right panels show the
data of the 47~Tuc cluster, observed with ground layer correction.
For comparison the position precision calculated from photon statistics as
the median of all frames is also shown ($\lozenge$).}
\label{fig:astr_position}
\end{figure*}

\subsection{Mean Positions}
As a last step we calculated the mean position for each star over
all corrected frames and its standard deviation as a measure of astrometric
precision. In Fig.~\ref{fig:astr_position} the achieved
astrometric precision is plotted over the K magnitude for each
star in the final lists of both data sets (blue $\times$ ). The given magnitude
represents the estimated 2MASS \citep{Skrutskie2006} K magnitude
of the stars, which suffices
to see the principal relation between precision and intensity. For
completeness the total counts are indicated at the upper $x$-axis
of the plots. These values show that we compare stars within 
the same detected flux range, even though the magnitude ranges differ. 
This is due to the fact that in the case of the observations of the 
cluster 47~Tuc the $Br_{\gamma}$ narrow-band filter was used instead of 
the broader $K_{s}$-filter and the exposure times were reduced compared 
to the observations of NGC~6388.\\ 

\noindent As one can see in the plots in Fig.~\ref{fig:astr_position}
for the NGC~6388 data set
(left panels), the fainter stars have less precision in their
position than the brighter ones. The mean positional precision of
the stars between 14 and 18 mag after the full distortion correction 
is 0.041~pixel corresponding to 1.143~mas in the
$x$-direction and 0.046~px (
1.278~mas) in the $y$-direction, where $x$ is parallel to the
right ascension and $y$ to the declination axes. The median value of the
precision in this magnitude range is slightly smaller, $x$:~0.039~px~(
1.084~mas),
$y$:~ 0.042~px~(1.179~mas). This is the
achievable astrometric precision with the available MAD data in
full MCAO mode.
Theoretically, as stated above in
\S~\ref{sec:residuals}, the faintest stars in this regime should
have a precision of about 0.015~px (0.420 mas) assuming
only photon statistics and then the mean positional precision in the same
magnitude range of K = 14 - 18~mag is 0.009~px (0.252 mas) in the
$x$- and $y$-direction.
The mean precision from photon statistics was thereby calculated by taking
the median of the positional precisions of each star in all individual frames.
These estimates for each star are also shown in Fig.~\ref{fig:astr_position},
plotted as diamonds. The measured precision of the mean position is a factor of
4.6
($x$-direction) and 5.1 ($y$-direction) worse than the one estimated from
photon statistics. This is a quite
large discrepancy, although one has to take more than simple photon
statistics into account for calculating a correct error budget, as for
example the error from the PSF estimation which was used to fit the stellar
positions and to calculate the uncertainty estimates. \\
\noindent In the GLAO data set of 47~Tuc the astrometric precision
for stars with corresponding 2MASS magnitudes between 9 and 12 mag
is 0.034~pixel (0.960~mas) and 0.035~pixel (0.972~mas) in the $x$- and $y$ directions, respectively. The median value
is 0.027~px (0.750~mas) in $x$ and 0.025~px (0.699~mas)
in $y$.
Although the fainter stars seem to have slightly larger
uncertainties, this correlation is less distinctive than in the
MCAO case (NGC~6388).
For comparison, the results from the basic distortion correction are also plotted
in Fig.~\ref{fig:astr_position} as black crosses.\\

\noindent In Tab.~\ref{tab:MAD_results} the above described results 
for the higher order correction are summarized.\\

\noindent Comparing the results for the ground layer correction
with those of the MCAO correction shows a higher precision in the
GLAO data. One would expect it the other way round as the initial
observing conditions and the average Strehl are better in the MCAO
data plus the MCAO correction is expected to correct the
wavefront distortions more accurately. Also the FWHM and the
diameter of 50\% ensquared energy are smaller in the MCAO data.
One of the
main differences in the two data sets is the jitter movement. As
already shown, this movement introduces distortions.\\ 
\noindent In a first
attempt we corrected only for shift, scale and rotation, but the afterwards
achieved precision is worse than expected, indicating distortions of higher
order. Correcting both data sets also for higher order distortions leads
to a higher precision on both, the NGC~6388 and the 47~Tuc data, but still
a better correction in the pure ground-layer correction can be seen (which is also
observed without jitter movements).

\begin{table*}
\vspace{0.3cm}
\caption{Summary of the expected and achieved astrometric
precisions after a distortion correction including higher orders.}
\label{tab:MAD_results}
\begin{center}
\begin{tabular}{|l|l|l|l|l|lDl|l|l|l|l|}
% \begin{tabular}{|l|l|l|l|l|l|lll|l|l|l|l|l|}
\hline
& \multicolumn{5}{cD}{NGC~6388, $K$ = 14-18} & \multicolumn{5}{c|}{47~Tuc, $K$ = 9-12} \\
unit & \multicolumn{2}{c|}{mean} & \multicolumn{2}{c|}{median} & Photon
& \multicolumn{2}{c|}{mean} & \multicolumn{2}{c|}{median} & Photon\\
& ~~~~$x$ & ~~~~$y$ & ~~~~~$x$ & ~~~~~$y$ & statistics  & ~~~~$x$ & ~~~~$y$ & ~~~~~$x$ & ~~~~~$y$ &
statistics  \\
\hline
px: & $\pm 0.041$ & $\pm 0.046$ & $\pm 0.039$ & $\pm 0.042$ & $\pm 0.009$ & $\pm
0.034$ & $\pm 0.035$ & $\pm 0.027$ & $\pm 0.025$ & $\pm 0.005$ \\
mas: & $\pm 1.143$ & $\pm 1.278$ & $\pm 1.084$ & $\pm 1.179$ & $\pm 0.252$ &
$\pm 0.960$ & $\pm 0.972$ & $\pm 0.750$ & $\pm 0.699$ & $\pm 0.084$ \\
\hline
\end{tabular}
\end{center}
% \caption{Summary of the expected and achieved astrometric
% precisions after a distortion correction including higher orders.
%  The first two columns for each data set
% (NGC~6388 and 47~Tuc) list the measured mean
% astrometric precision and the next two columns the median
% values of the precision, measured for stars in the indicated
% magnitude ranges. The sixth and eleventh column list the expected
% precision from photon statistics for the two data sets.}
\end{table*}

% %______________________________________________________________
%
 \section{Conclusions}
We have analyzed the first multi conjugated adaptive optics data
available in the layer oriented approach with respect to
astrometric performance. The data were taken with the MCAO
demonstrator MAD at the VLT. Two sets of data of globular
clusters, observed in two different approaches were analyzed: the
globular cluster 47~Tucanae with ground layer correction only and
the globular cluster NGC~6388 in full two-layer MCAO correction.\\

\noindent As a performance measure we calculated
Strehl maps for each frame. The Strehl is fairly uniform over the
FoV with a small degradation toward the edges of the FoV and average
values between 11\% and 23\% in the MCAO data and between 9\% and 14\% in the GLAO data.
 The lower Strehl in the 47~Tuc data set may partly be
explained by the fact that only the distortions due to the ground
layer were corrected, but also the initial atmospheric conditions
were worse.\\
After extensive PSF tests we analyzed the data with the {\it
StarFinder} code. We created a master frame with positions of
isolated stars in the field and calculated in a first attempt
distortion parameters
for shift and scale in $x-$ and $y$-direction and a rotation for
each frame to this master frame. Separation measurements between
stars before and after the distortion correction showed that these
corrections are indeed reducing the scatter in the separations
measured over all frames (\S~\ref{subsec:distMeas}). But it also
shows a residual scatter, which is probably due to higher order
distortions. A pattern visible in the separation measurements
(Fig.~\ref{fig:separation}) as well as in the applied distortion
parameters (Fig.~\ref{fig:distortions}) is thought to be due to
the jitter movement of the camera during the observations. This
movement introduced additional distortions which could only be
corrected partly, with this distortion correction. To exploit the full
capacity of astrometry with MCAO we performed a $4^{th}$ order polynomial
distortion correction, including also higher order terms. \\
\noindent The mean precision of the positions of the stars, calculated
by the scatter of the mean position of the stars over all frames,
is 0.041~pixel (1.143~mas) and 0.046~pixel
(1.278~mas), for the $x$- and $y$-direction, respectively, in the
corresponding 2MASS K magnitude range from 14 to 18 mag in the
NGC~6388 data set (MCAO). In the 47~Tuc data set (GLAO) the mean
precision is 0.034~pixel (0.9602~mas) and 0.035~pixel (0.972~mas) for 
comparable $K$ magnitudes between 9 and
12.\\
These results show impressively the capacity of high precision
astrometry over a large field of view observed with MCAO.\\

\noindent An astrometric analysis of the core of 47~Tuc was also
performed by \cite{McLaughlin47Tuc}, who used several epochs of
data from the Hubble Space Telescope (HST). They derived
positional precisions in the single epoch data, for stars in the
same area as the here analyzed FoV, taken with the High Resolution
Camera (HRC) of the Advanced Camera for Surveys (ACS) for most
stars in the range of 0.01-0.05 pixel. With a plate-scale of 0.027
arcsec/pixel this corresponds to 0.27-1.35 mas. The errors were
calculated in the same way as in this work, taking the standard
deviation of the positions in all frames as uncertainties.
Detailed distortion corrections were computed for ACS by
\cite{Anderson2002}, which were applied to the data in the work of
McLaughlin et al.. This shows that the precision derived with MAD
is already comparable to HST/ACS astrometry and with a good distortion
characterization, future instruments could yield even
higher astrometric precision.\\

\noindent Although the Strehl-ratio is smaller and the FWHM is larger in
the GLAO data of the cluster 47~Tuc, the achieved astrometric
precision is higher. Also the observing conditions were worse
during the GLAO observations compared to the MCAO observations
with a mean seeing of $1.13\arcsec$ and $0.46\arcsec$,
respectively. All this leads to the conclusion that the
degradation of the astrometric precision in the MCAO data set is
mainly due to the jitter movement during the observations, which
introduced additional distortions. But also the more complex
correction of two layers could have introduced distortions, which we could not
correct for. To fully characterize
the remaining distortions, one would need to analyse more data,
taken under various seeing conditions and observation
configurations. As MAD will not be offered again, a fully
satisfactory analysis is not possible at this point. Nevertheless
one can interpret the remaining positional uncertainty in the GLAO
corrected data, which was obtained without any jitter movement, as
distortions remaining from the AO correction and not
compensated turbulence.\\

\noindent All the results presented here are still given in
detector coordinates, as we analyzed the data in matters of the
adaptive optics correction and instrumentation stability over the
time of the full length of the observation. Going to celestial
coordinates would involve the correction for effects such as
differential aberration and differential refraction to derive the
true positions of the stars. As the observed FoV is large, these
effects can reach several milliseconds of arc of displacement
between stars at different points on the detector \cite[][in
preparation]{Meyer2010GJ}. These transformations introduce
additional position uncertainties, degrading the astrometric
precision further, but need to be performed when comparing data
from different epochs, as for example in proper motion studies.
The data analyzed here is single epoch data, therefore these
corrections did not need to be performed to investigate the
stability and possible accuracy of astrometric measurements
in MCAO data, as these are effects present in all ground-based
imaging data and do not influence or are influenced by the AO
performance.\\
To make a final comparison between ground-based MCAO and space-based 
astrometric precision, a multi-epoch study needs to be carried out. 
As MAD is not offered again, this is not possible at the current 
state and with the available data.

\begin{acknowledgements}
The data analyzed here are based on observations collected at the European
Southern Observatory, Paranal, Chile, as part of the MAD Guaranteed
Time Observations. We would like to thank the anonymous referee for the useful comments and suggestions to this paper.
\end{acknowledgements}

% % \bibliography{/home/meyer/Documents/papers/BibTeX/Literatur}
\bibliography{/data1/MPIA/Documents/papers/BibTeX/Literatur}

\small
\begin{thebibliography}{33}
\expandafter\ifx\csname natexlab\endcsname\relax\def\natexlab#1{#1}\fi

\bibitem[{{Anderson}(2002)}]{Anderson2002}
{Anderson}, J. 2002, in The 2002 HST Calibration Workshop : Hubble after the
  Installation of the ACS and the NICMOS Cooling System, ed. {S.~Arribas,
  A.~Koekemoer, \& B.~Whitmore}, 13--+

\bibitem[{{Arcidiacono} {et~al.}(2006){Arcidiacono}, {Lombini}, {Diolaiti},
  {Farinato}, \& {Ragazzoni}}]{Arcidiacono2006}
{Arcidiacono}, C., {Lombini}, M., {Diolaiti}, E., {Farinato}, J., \&
  {Ragazzoni}, R. 2006, in Society of Photo-Optical Instrumentation Engineers
  (SPIE) Conference Series, Vol. 6272, 627227

\bibitem[{{Arcidiacono} {et~al.}(2010){Arcidiacono}, {Lombini}, {Moretti},
  {Ragazzoni}, {Farinato}, {Falomo}, {Gullieuszik}, \&
  {Piotto}}]{Arcidiacono2010}
{Arcidiacono}, C., {Lombini}, M., {Moretti}, A., {et~al.} 2010, in Society of
  Photo-Optical Instrumentation Engineers (SPIE) Conference Series, Vol. 7736,
  Society of Photo-Optical Instrumentation Engineers (SPIE) Conference Series

\bibitem[{{Arcidiacono} {et~al.}(2008){Arcidiacono}, {Lombini}, {Ragazzoni},
  {Farinato}, {Diolaiti}, {Baruffolo}, {Bagnara}, {Gentile}, {Schreiber},
  {Marchetti}, {Kolb}, {Tordo}, {Donaldson}, {Soenke}, {Oberti}, {Fedrigo},
  {Vernet}, \& {Hubin}}]{Arcidiacono2008}
{Arcidiacono}, C., {Lombini}, M., {Ragazzoni}, R., {et~al.} 2008, in Society of
  Photo-Optical Instrumentation Engineers (SPIE) Conference Series, Vol. 7015,
  Society of Photo-Optical Instrumentation Engineers (SPIE) Conference Series

\bibitem[{{Bean} {et~al.}(2007){Bean}, {McArthur}, {Benedict}, {Harrison},
  {Bizyaev}, {Nelan}, \& {Smith}}]{Bean2007}
{Bean}, J.~L., {McArthur}, B.~E., {Benedict}, G.~F., {et~al.} 2007, \aj, 134,
  749

\bibitem[{Beckers(1988)}]{Beckers1988}
Beckers, J. 1988, in Very Large Telescopes and their Instrumentation, ESO
  Conference and Workshop Proceedings, p.693, Garching, March 21-24, 1988,
  edited by Marie-Helene Ulrich., 693

\bibitem[{{Benedict} {et~al.}(2002){Benedict}, {McArthur}, {Forveille},
  {Delfosse}, {Nelan}, {Butler}, {Spiesman}, {Marcy}, {Goldman}, {Perrier},
  {Jefferys}, \& {Mayor}}]{Benedict2002}
{Benedict}, G.~F., {McArthur}, B.~E., {Forveille}, T., {et~al.} 2002, \apjl,
  581, L115

\bibitem[{{Cresci} {et~al.}(2005){Cresci}, {Davies}, {Baker}, \&
  {Lehnert}}]{Cresci2005}
{Cresci}, G., {Davies}, R.~I., {Baker}, A.~J., \& {Lehnert}, M.~D. 2005, \aap,
  438, 757

\bibitem[{{Diolaiti} {et~al.}(2000{\natexlab{a}}){Diolaiti}, {Bendinelli},
  {Bonaccini}, {Close}, {Currie}, \& {Parmeggiani}}]{Diolaiti20002}
{Diolaiti}, E., {Bendinelli}, O., {Bonaccini}, D., {et~al.} 2000{\natexlab{a}},
  147, 335

\bibitem[{{Diolaiti} {et~al.}(2000{\natexlab{b}}){Diolaiti}, {Bendinelli},
  {Bonaccini}, {Close}, {Currie}, \& {Parmeggiani}}]{Diolaiti2000}
{Diolaiti}, E., {Bendinelli}, O., {Bonaccini}, D., {et~al.} 2000{\natexlab{b}},
  in Presented at the Society of Photo-Optical Instrumentation Engineers (SPIE)
  Conference, Vol. 4007, Society of Photo-Optical Instrumentation Engineers
  (SPIE) Conference Series, ed. {P.~L.~Wizinowich}, 879--888

\bibitem[{{Ellerbroek} {et~al.}(1994){Ellerbroek}, {van Loan}, {Pitsianis}, \&
  {Plemmons}}]{Ellerbroeck1994}
{Ellerbroek}, B.~L., {van Loan}, C., {Pitsianis}, N.~P., \& {Plemmons}, R.~J.
  1994, in Society of Photo-Optical Instrumentation Engineers (SPIE) Conference
  Series, Vol. 2201, Society of Photo-Optical Instrumentation Engineers (SPIE)
  Conference Series, ed. {M.~A.~Ealey \& F.~Merkle}, 935--948

\bibitem[{{Farinato} {et~al.}(2008){Farinato}, {Ragazzoni}, {Arcidiacono},
  {Brunelli}, {Dima}, {Gentile}, {Viotto}, {Diolaiti}, {Foppiani}, {Lombini},
  {Schreiber}, {Bizenberger}, {De Bonis}, {Egner}, {G{\"a}ssler}, {Herbst},
  {K{\"u}rster}, {Mohr}, \& {Rohloff}}]{Farinato2008}
{Farinato}, J., {Ragazzoni}, R., {Arcidiacono}, C., {et~al.} 2008, in Society
  of Photo-Optical Instrumentation Engineers (SPIE) Conference Series, Vol.
  7015, Society of Photo-Optical Instrumentation Engineers (SPIE) Conference
  Series

\bibitem[{{Fritz} {et~al.}(2010){Fritz}, {Gillessen}, {Trippe}, {Ott},
  {Bartko}, {Pfuhl}, {Dodds-Eden}, {Davies}, {Eisenhauer}, \&
  {Genzel}}]{Fritz2010}
{Fritz}, T., {Gillessen}, S., {Trippe}, S., {et~al.} 2010, \mnras, 401, 1177

\bibitem[{{Hubin} {et~al.}(2002){Hubin}, {Marchetti}, {Fedrigo}, {Conan},
  {Ragazzoni}, {Diolaiti}, {Tordi}, {Rousset}, {Fusco}, {Madec}, {Butler},
  {Stefan}, \& {Esposito}}]{Hubin2002}
{Hubin}, N., {Marchetti}, E., {Fedrigo}, E., {et~al.} 2002, in European
  Southern Observatory Astrophysics Symposia, Vol.~58, European Southern
  Observatory Astrophysics Symposia, ed. {E.~Vernet, R.~Ragazzoni, S.~Esposito,
  \& N.~Hubin}, 27--+

\bibitem[{{King} \& {Anderson}(2001)}]{King2001}
{King}, I.~R. \& {Anderson}, J. 2001, in Astronomical Society of the Pacific
  Conference Series, Vol. 228, Dynamics of Star Clusters and the Milky Way, ed.
  {S.~Deiters, B.~Fuchs, A.~Just, R.~Spurzem, \& R.~Wielen}, 19--+

\bibitem[{{Marchetti} {et~al.}(2007){Marchetti}, {Brast}, {Delabre},
  {Donaldson}, {Fedrigo}, {Frank}, {Hubin}, {Kolb}, {Lizon}, {Marchesi},
  {Oberti}, {Reiss}, {Santos}, {Soenke}, {Tordo}, {Baruffolo}, {Bagnara}, \&
  {The CAMCAO Consortium}}]{Marchetti2007}
{Marchetti}, E., {Brast}, R., {Delabre}, B., {et~al.} 2007, The Messenger, 129,
  8

\bibitem[{{Marchetti} {et~al.}(2003){Marchetti}, {Hubin}, {Fedrigo}, {Brynnel},
  {Delabre}, {Donaldson}, {Franza}, {Conan}, {Le Louarn}, {Cavadore},
  {Balestra}, {Baade}, {Lizon}, {Gilmozzi}, {Monnet}, {Ragazzoni},
  {Arcidiacono}, {Baruffolo}, {Diolaiti}, {Farinato}, {Vernet-Viard}, {Butler},
  {Hippler}, \& {Amorin}}]{Marchetti2003}
{Marchetti}, E., {Hubin}, N.~N., {Fedrigo}, E., {et~al.} 2003, in Society of
  Photo-Optical Instrumentation Engineers (SPIE) Conference Series, Vol. 4839,
  Society of Photo-Optical Instrumentation Engineers (SPIE) Conference Series,
  ed. {P.~L.~Wizinowich \& D.~Bonaccini}, 317--328

\bibitem[{{Markwardt}(2009)}]{Markwardt2009}
{Markwardt}, C.~B. 2009, in Astronomical Society of the Pacific Conference
  Series, Vol. 411, Astronomical Society of the Pacific Conference Series, ed.
  {D.~A.~Bohlender, D.~Durand, \& P.~Dowler}, 251--+

\bibitem[{{McLaughlin} {et~al.}(2006){McLaughlin}, {Anderson}, {Meylan},
  {Gebhardt}, {Pryor}, {Minniti}, \& {Phinney}}]{McLaughlin47Tuc}
{McLaughlin}, D.~E., {Anderson}, J., {Meylan}, G., {et~al.} 2006, \apjs, 166,
  249

\bibitem[{{Meyer} {et~al.}(2010){Meyer}, {K{\"u}rster}, \&
  {K{\"o}hler}}]{Meyer2010GJ}
{Meyer}, E., {K{\"u}rster}, M., \& {K{\"o}hler}, R. 2010, in preparation

\bibitem[{{Moretti} {et~al.}(2009){Moretti}, {Piotto}, {Arcidiacono}, {Milone},
  {Ragazzoni}, {Falomo}, {Farinato}, {Bedin}, {Anderson}, {Sarajedini},
  {Baruffolo}, {Diolaiti}, {Lombini}, {Brast}, {Donaldson}, {Kolb},
  {Marchetti}, \& {Tordo}}]{Moretti2009}
{Moretti}, A., {Piotto}, G., {Arcidiacono}, C., {et~al.} 2009, \aap, 493, 539

\bibitem[{Ragazzoni(1996)}]{Ragazzoni1996}
Ragazzoni, R. 1996, J. Modern Opt., \textbf{43}, 289

\bibitem[{{Ragazzoni} {et~al.}(2000{\natexlab{a}}){Ragazzoni}, {Farinato}, \&
  {Marchetti}}]{Ragazzoni2000b}
{Ragazzoni}, R., {Farinato}, J., \& {Marchetti}, E. 2000{\natexlab{a}}, in
  Presented at the Society of Photo-Optical Instrumentation Engineers (SPIE)
  Conference, Vol. 4007, Society of Photo-Optical Instrumentation Engineers
  (SPIE) Conference Series, ed. {P.~L.~Wizinowich}, 1076--1087

\bibitem[{{Ragazzoni} {et~al.}(2000{\natexlab{b}}){Ragazzoni}, {Marchetti}, \&
  {Valente}}]{Ragazzoni2000}
{Ragazzoni}, R., {Marchetti}, E., \& {Valente}, G. 2000{\natexlab{b}}, \nat,
  403, 54

\bibitem[{{Rigaut}(2002)}]{Rigaut2002}
{Rigaut}, F. 2002, in Beyond conventional adaptive optics : a conference
  devoted to the development of adaptive optics for extremely large telescopes.
  Proceedings of the Topical Meeting held May 7-10, 2001, Venice, Italy. Edited
  by E. Vernet, R. Ragazzoni, S. Esposito, and N. Hubin. Garching, Germany:
  European Southern Observatory, 2002 ESO Conference and Workshop Proceedings,
  Vol. 58, ISBN 3923524617, p.11

\bibitem[{{Rochau} {et~al.}(2010){Rochau}, {Brandner}, {Stolte}, {Gennaro},
  {Gouliermis}, {Da Rio}, {Dzyurkevich}, \& {Henning}}]{Rochau2010}
{Rochau}, B., {Brandner}, W., {Stolte}, A., {et~al.} 2010, \apjl, 716

\bibitem[{Roddier(1999)}]{Roddier1999}
Roddier, F. 1999, \textit{Adaptive Optics in Astronomy} (Cambridge University
  Press)

\bibitem[{{Sch{\"o}del}(2010)}]{Schoedel2010}
{Sch{\"o}del}, R. 2010, \aap, 509, A58+

\bibitem[{{Sch{\"o}del} {et~al.}(2009){Sch{\"o}del}, {Merritt}, \&
  {Eckart}}]{Schoedel2009}
{Sch{\"o}del}, R., {Merritt}, D., \& {Eckart}, A. 2009, \aap, 502, 91

\bibitem[{{Skrutskie} {et~al.}(2006){Skrutskie}, {Cutri}, {Stiening},
  {Weinberg}, {Schneider}, {Carpenter}, {Beichman}, {Capps}, {Chester},
  {Elias}, {Huchra}, {Liebert}, {Lonsdale}, {Monet}, {Price}, {Seitzer},
  {Jarrett}, {Kirkpatrick}, {Gizis}, {Howard}, {Evans}, {Fowler}, {Fullmer},
  {Hurt}, {Light}, {Kopan}, {Marsh}, {McCallon}, {Tam}, {Van Dyk}, \&
  {Wheelock}}]{Skrutskie2006}
{Skrutskie}, M.~F., {Cutri}, R.~M., {Stiening}, R., {et~al.} 2006, \aj, 131,
  1163

\bibitem[{{Tallon} \& {Foy}(1990)}]{Tallon1990}
{Tallon}, M. \& {Foy}, R. 1990, \aap, 235, 549

\bibitem[{{Trippe} {et~al.}(2008){Trippe}, {Gillessen}, {Gerhard}, {Bartko},
  {Fritz}, {Maness}, {Eisenhauer}, {Martins}, {Ott}, {Dodds-Eden}, \&
  {Genzel}}]{Trippe2008}
{Trippe}, S., {Gillessen}, S., {Gerhard}, O.~E., {et~al.} 2008, \aap, 492, 419

\bibitem[{{Yan} {et~al.}(2005){Yan}, {Wu}, {Li}, \& {Chen}}]{Yan2005}
{Yan}, H., {Wu}, H., {Li}, S., \& {Chen}, S. 2005, in Society of Photo-Optical
  Instrumentation Engineers (SPIE) Conference Series, Vol. 5903, Society of
  Photo-Optical Instrumentation Engineers (SPIE) Conference Series, ed.
  {R.~K.~Tyson \& M.~Lloyd-Hart}, 260--271

\end{thebibliography}
% % \bibliography{Literatur}
% %

%\small

%\bibliographystyle{}
% \bibliography{mad_paper_accepted_pdf}
% \begin{thebibliography}{33}
% \end{thebibliography}

 \end{document}